\documentclass[aps,pra,showpacs,twoside,twocolumn,10pt]{revtex4-1}
\usepackage[colorlinks=true, citecolor=red, urlcolor=blue ]{hyperref}
\usepackage{epsfig,newlfont,amssymb,amsfonts,amsmath,bm,subfigure,palatino,mathtools,amsthm,braket,times,soul,enumitem,color}
\usepackage[normalem]{ulem}
\newcommand{\stkout}[1]{\ifmmode\text{\sout{\ensuremath{#1}}}\else\sout{#1}\fi}
\definecolor{magenta}{rgb}{1.0, 0.0, 0.56}

\begin{document}
 
 %Regional versus global heat currents in quantum thermal devices
 
 %\title{Local versus global heat currents and entropy production rates in quantum thermal devices}
 
% \title{Local balance relation between entropy production rate and heat current\\in non-equilibrium quantum devices}
 
%\title{Local EPR relation and bound of total heat current deficit}

%\title{Multiparty Spohn's relation for mixed local Markovian and non-Markovian quantum dynamics}

\title{Multiparty Spohn's theorem\\for a combination of local Markovian and non-Markovian quantum dynamics}

\author{Ahana Ghoshal and Ujjwal Sen}
\affiliation{Harish-Chandra Research Institute,  A CI of Homi Bhabha National Institute, Chhatnag Road, Jhunsi, Prayagraj 211 019, India}

\begin{abstract}
We obtain a Gorini-Kossakowski-Sudarshan-Lindblad -like master equation for two or more quantum systems connected locally to a combination of Markovian and non-Markovian heat baths. The master equation was originally formulated for multiparty systems with either exclusively Markovian or non-Markovian environments. We extend it to encompass the case of multiple quantum systems connected to a mixture of Markovian and non-Markovian heat baths. The coexistence of both non-Markovian and Markovian environments is a plausible scenario, particularly when studying hybrid physical systems such as atom-photon arrangements.
%
%\textcolor{red}{\sout{This  broadening of the area of investigation of open quantum dynamics is also realistic, as a combination of non-Markovian and Markovian environments is a reasonable possibility, especially when considering hybrid physical systems like atom-photon arrangements.}}
%
We analyze the thermodynamic quantities for such a set of local environments, and derive a modified form of the Spohn's theorem for the setup. The modification of the theorem naturally leads to a witness as well as an easily computable quantifier of non-Markovianity. 
%in some particular situations, in which the local canonnical equilibrium states of the composite system do not act as the stationary  states of the partial superoperators of the total Lindblad master equation in combined local Markovian and non-Markovian dynamics. Furthermore, we observe that for a combined local Markovian and non-Markovian evolution, in the initial time span the non-Markovian baths have a significant contribution,
Expectedly, we find that for multiparty situations, where a combination of Markovian and non-Markovian heat baths are active, the response in thermodynamic system characteristics due to non-Markovian baths is prominent at times close to the initial time of evolution, whereas the long-time behavior is predominantly controlled by the Markovian ones.
\end{abstract}

\maketitle
\noindent
\section{Introduction}
%\emph{\textbf{Introduction.--}}
Quantum thermodynamics is an emerging field of research and the interconnections of it with quantum information theory have been studied from myriad perspectives. 
%The overlapping research area of quantum thermodynamics and quantum information 
%\textcolor{red}{\sout{The interface has been enriched with the inventions of quantum thermal devices~\cite{Palao,Feldmann,Levy1,Levy,Uzdin,Clivaz,Mitchison,Yuan,Joulain,Zhang,Su,Mandarino}. The dynamics of quantum thermal machines are governed by the theory of open quantum systems~\cite{Petruccione,Alicki,Rivas,Lidar}, and the investigations, in particular, have led to a better understanding of the thermodynamic laws in the quantum regime~\cite{Allahverdyan,Gemmer,Kosloff,Brand,Gardas,Gelbwaser,Misra,Millen,Vinjanampathy,Goold,Benenti,Binder,Deffner}.}} 
The study of quantum thermal devices~\cite{Palao,Feldmann,Levy1,Levy,Uzdin,Clivaz,Mitchison,Yuan,Joulain,Zhang,Su,Mandarino} and that of their dynamics governed by open quantum systems~\cite{Petruccione,Alicki,Rivas,Lidar}, has significantly contributed to the understanding of thermodynamics in the quantum regime~\cite{Allahverdyan,Gemmer,Kosloff,Brand,Gardas,Gelbwaser,Misra,Millen,Vinjanampathy,Goold,Benenti,Binder,Deffner}. 
%While a substantial body of work analyzes quantum devices under Markovian evolution, non-Markovian dynamics has also been explored.} 
A significant body of work that analyze quantum devices deals with Markovian evolution, although non-Markovian dynamics has also been considered. Markovian environments are rare in nature and they exhibit rather specific behaviors~\cite{Petruccione,Kampen}. The bosonic baths with infinite numbers of harmonic oscillators, within some restrictions, usually behave as Markovian environments, while certain thermal baths, such as spin baths~\cite{Prokof'ev, Fischer, Camalet, Bhattacharya, Majumdar,Breuer1,Hutton}, do not fit the Markovian framework easily and are categorized as non-Markovian reservoirs. Some non-Markovian baths may have Markovian limits, but for systems such as the spin star model, this limit can be elusive~\cite{Breuer1}. Detecting and characterizing non-Markovianity has been achieved through various measures~\cite{Plenio1,Piilo,Huelga,Rivas1,Zeng1,Debarba,Strasberg,Das,Huang}, which are not all equivalent.
%The basic principles of the quantum thermal machines are based on the interplay between the thermodynamic laws in the quantum regimes, which has been introduced and investigated in~\cite{Allahverdyan,Gemmer,Kosloff,Brand,Gardas,Gelbwaser,Misra,Millen,Vinjanampathy,Goold,Benenti,Binder,Deffner}, and the theory of open quantum systems~\cite{Petruccione,Alicki,Rivas,Lidar}. An extensive area in open quantum systems is developed on the Lindblad master equations which is constructed under some certain assumptions called the Born-Markov assumptions. The evolution of the systems considering these approximations is called the Markovian evolution and the violation of these assumptions leads to an evolution which is non-Markovian.\par
%The detection of an evolution to be a Markovian or a non-Markovian one depends on the 
\par
%Two widely used measures are those proposed by Breuer-Laine-Pillo (BLP)~\cite{Piilo} and Rivas-Huelga-Plenio (RHP)~\cite{Huelga}, which respectively uses non-moreuer, E.-M. Laine, and J. Piilo, notonicity in time-evolution of  state distinguishability and entanglement.
%The mostly used non-Markovianity measures are proposed by Breuer et al. (BLP measure)~\cite{Piilo} and Rivas et al. (RHP measure)~\cite{Huelga}. 
%\textcolor{red}{The BLP measure detects the distinguishability between any two states in time, a growth of which indicates information backflow from the environment to the open system. The growth of distinguishability of any two states is caused by a non-divisible map, and as a consequence, a necessary, but not sufficient, condition turns out that a nondivisible dynamical map represents a non-Markovian evolution. On the other hand according to RHP measure a complete positive trace preserving (CPTP) map describes a Markovian evolution.} 
%Later on, an equivalence of these two measures in some cases has been established~\cite{Rivas1,Zeng1}. %After the BLP and RHP approaches of non-Markovianity measure, some similar measures and quantifiers of non-Markovianity has been suggested and the equivalence between the different measures has been studied.
%See e.g.~\cite{Debarba,Strasberg,Das,Huang} for some further works on non-Markovianity. \par
Heat current and entropy production rate (EPR) are two fundamental quantities that give an idea about the thermal properties of a system. 
%\textcolor{red}{
The second law of thermodynamics leads to a balance equation, relating EPR ($\sigma$), the von Neumann entropy ($S$), and heat current ($J$) for a single system immersed in a heat bath, given by $\sigma=\frac{dS}{dt} + J$.
%and 
%the validation of all applications of quantum thermodynamics can be described by this equation.} 
%The quantity, EPR, is not only important in describing the quantum thermodynamics, its contribution to detect the non-Markovianity of an evolution is also significant. 
%\textcolor{blue}{\sout{EPR is an important characteristic for understanding the thermodynamics of a system, and moreover, is a valuable physical quantity for detection of non-Markovianity of a dynamics.}} 
Spohn's theorem~\cite{Spohn,Lebowitz} states that for a Markovian evolution, with bath initial states being thermal, EPR of the system is always positive. It is known that for non-Markovian evolutions, the EPR may take negative values~\cite{Bylicka,Marcantoni,Bhattacharya,Popovic,Strasberg1,Naze,Bonanca,Ghoshal}.
%So, that EPR can be treated as a non-Markovianity detector. While speaking about detectors and quantifiers of non-Markovianity, we remember that many of them provide either a necessary or a sufficient criterion for non-Markovianity detection, but not both. 
For a deeper understanding of the entropy production rate, see, e.g.,~\cite{Alicki1,Turitsyn,Andrae,Vollmer,Breuer,Esposito,Abe,Bandi,Deffner1,Gardas,Salis,Carlen,Xu1,Kawazura,Hase,Kuzovlev,Banerjee,Yu,Xu,Wang,Chen,Taye,Zhang1,Ploskon,Jaramillo,Santos,Zeng,Busiello,Dixit,Kanda,Seara,Lee,Wolpert,Goes,Przymus,Budhiraja,Zicari,Kappler,Gibbins,Li,Horowitz,Tietz,Jiang1,Rossi}.\par

In physical systems, the presence of memory effects and strong system-bath correlations may lead to deviations from Markovian dynamics. In some cases, certain components or interactions within a system may exhibit Markovian behavior, while others display non-Markovian behavior. This can arise due to the complexity of the system's architecture or the interplay between different timescales involved in the dynamics. 
%\textcolor{blue}{\sout{So, the study of combined Markovian and non-Markovian evolution of a system is necessary. Furthermore, it is important to note that}}
For example, this type of model holds significant relevance as a plausible approach for investigating hybrid systems, such as atom-photon arrangements. In atom-photon systems~\cite{Wang_2016,Vural_2018,Fehler_2021}, the timescales of atomic and photonic interactions with their respective environments can vary. For instance, atomic transitions may occur on a different timescale compared to the relaxation processes involving emitted or absorbed photons. This can lead to a situation where certain aspects of the system-environment interaction exhibit Markovian characteristics, while others show non-Markovian features. Moreover, in atom-photon setups, the environment may not be homogeneous. 
%It could consist of different modes or channels with varying characteristics.
Some components of the environment may induce memory effects and correlations, resulting in a non-Markovian influence, while other components may exhibit Markovian behavior. So, understanding and characterizing the interplay between Markovian and non-Markovian elements in atom-photon systems is essential for optimizing their performance in quantum technologies and information processing applications.
\par
%Indeed, non-Markovian evolutions have been widely studied in literature along with Markovian ones, especially when all the relevant environments in the physical system considered are either non-Markovian or Markovian.\par
%even do not have any Markovian limit~\cite{Breuer1}. Therefore the non-Markovian approach in open system evolution is more practical. The global Markovian as well as non-Markovian evolutions has been broadly studied in the literature, but there may arise a complicated situation where we need to study a special type of dynamical evolution originating from a combination of local Markovian and non-Markovian dynamics.\par
Here, we consider a situation where the local parts of the physical system under study are affected by local environments, 
%\textcolor{blue}{some of them are non-Markovian and others are not so. We derive a Gorini-Kossakowski-Sudarshan-Lindblad (GKSL) -like master equation, analyze thermodynamic quantities like heat current and EPR for the composite system, and obtain a modified form of the well-known Spohn's theorem~\cite{Spohn,Lebowitz} for this multiparty setup. Furthermore, we} 
which can be a few non-Markovian and the remaining not so. We
%In our paper we consider a situation of local Markovian and non-Markovian combined evolution and 
derive a Gorini-Kossakowski-Sudarshan-Lindblad (GKSL) -like master equation of the system for this case 
%We first provide the analytical derivation for a two-qubit system and then extend it to the case of $m+n$ subsystems, locally connected to $m$ Markovian and $n$ non-Markovian baths respectively.
and study the thermodynamic quantities such as heat current and EPR, for the composite system. %\textcolor{red}{\sout{(of $m+n$ subsystems) immersed in a ``mixed" set of local environments and}} %\textcolor{blue}{under this circumstance}, 
Furthermore, we obtain a modified form of the well-known Spohn's theorem~\cite{Spohn,Lebowitz} in connection to the second law of thermodynamics for this multiparty setup and %Here and in the rest of the paper, by a ``mixed set of local environments or baths", we will mean a situation where a few subsystems of the system are engaged with non-Markovian baths, while the remaining are so with Markovian ones. 
%\textcolor{red}{\sout{Moreover, we}}
propose an easily computable quantity that 
%which is easily computable and 
can be treated as a quantifier of non-Markovianity. 
%\textcolor{red}{\sout{in case of a mixed set of local environments}}. 
%We explicitly consider the case of four system qubits, with each of the qubits being connected to a non-Markovian or a Markovian bath. 
%\textcolor{blue}{For a four-qubit system under the combined local Markovian and non-Markovian dynamics, we observe that, non-Markovian effects on thermodynamic system characteristics dominate early on, but diminish over time, leading to more Markovian-like behavior.}  
For a four-qubit system, under the combined evolution of local Markovian and non-Markovian dynamics we observe that, %\textcolor{red}{\sout{in a mixed set of local environments}}
the response in thermodynamic system characteristics is dominated by the effect of non-Markovian baths at short times. 
%local Markovian and non-Markovian combined evolution. Further, we demonstrate a simple example to get an idea about the nature of the quantifier for a four-qubit system and observe that for a very short time the non-Markovian baths dominate. 
However, as expected, with the increase of time, non-Markovianity effects reduce, and the dynamics is more and more Markovian-like.\par
%of the system reduces and the dynamics will prone to be more and more Markovian like. \par
%The remainder of the paper is arranged as follows. In Sec.~\ref{sec:2} we derive the Gorini-Kossakowski-Sudarshan-Lindblad master equation (GKSL equation) of the system for a mixed set of local environs, for $m+n$ qubits. Section~\ref{sec:3} provides definitions  of some physical quantities like heat current and EPR, which are used to describe the second law of thermodynamics in the quantum regime. In Sec.~\ref{sec:4}, we show that the Spohn's theorem has a modified form in case of a mixed set of local environments of which some are non-Markovian while others are Markovian. This leads to an extra term in the Spohn's relation, which we treat %Markovian and non-Markovian combined evolution and hence gives rise to an extra term in the general equation of the theorem which can be treated 
%as a quantifier of non-Markovianity. The nature of the quantifier for mixed environments is discussed in the same section. 
%and the behaviour of combined local Markovian and non-Markovian dynamics are also demonstrated in Sec.~\ref{sec:4}. 
%Section~\ref{sec:5} contains the concluding remarks.
\noindent
\section{Multiparty GKSL-like equation for local Markovian and non-Markovian baths}
\begin{figure} 
\includegraphics[height=5cm,width=8.7cm]{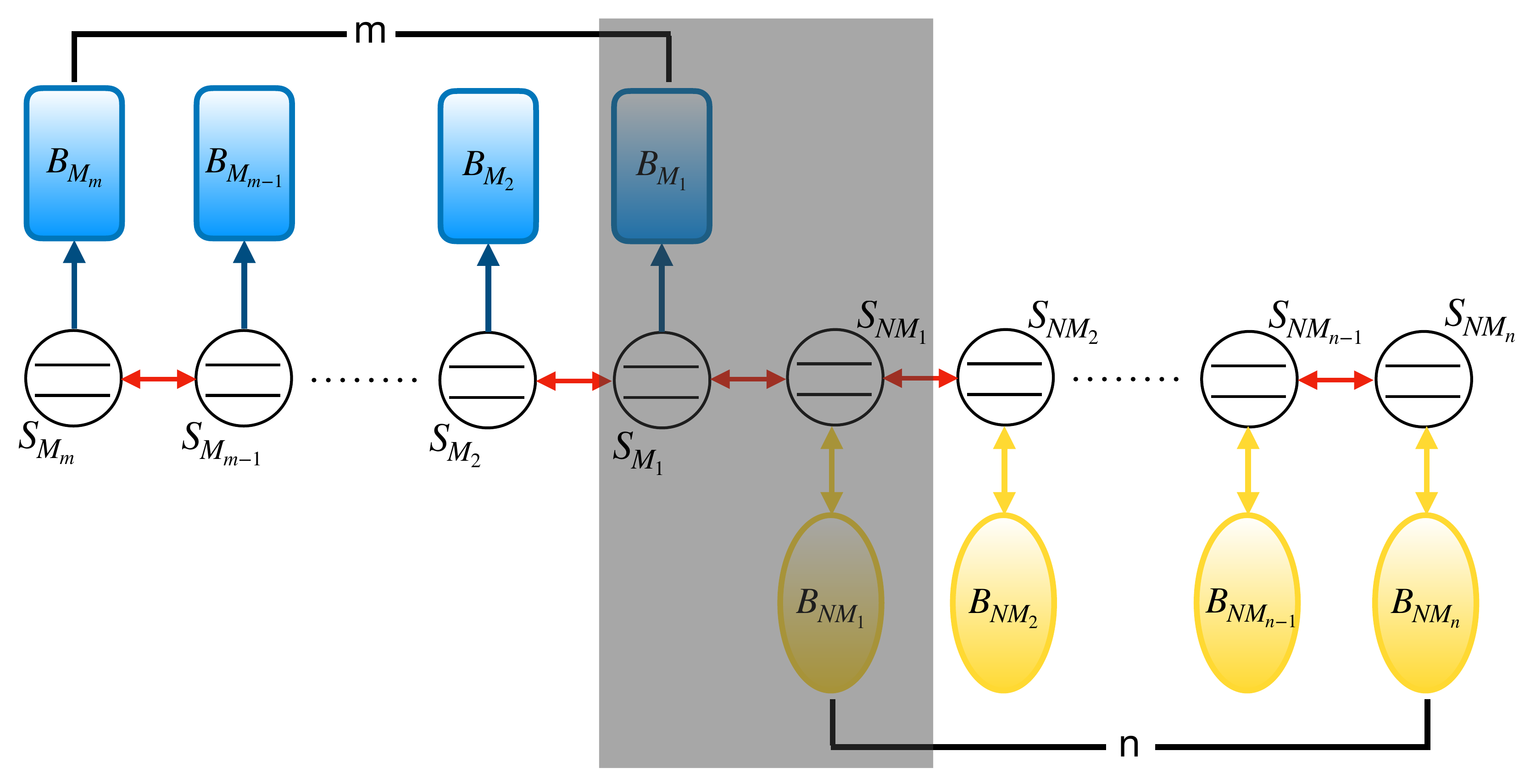}
\caption{
%\textcolor{red}{\sout{Mixed set of local environments.}} 
Combination of local Markovian and non-Markovian dynamics of a multiparty system.
%In the left panel, 
Here we present a schematic diagram of $m+n$ subsystems, evolving under their system Hamiltonian as well as local environments, some of which are Markovian and the rest are not so. $B_{M_1} \cdots B_{M_m}$ are the baths which can be treated under the Born-Markov approximations, hence are Markovian baths, and $B_{N_1} \cdots B_{N_n}$ are the baths residing in a non-Markovian family. A simplified scenario with only two subsystems interacting with two baths locally, among which one is Markovian and the other is non-Markovian is presented in the gray box.
%A special case of the left panel is presented in the right one where only two TLSs are interacting with two baths locally among which one is Markovian and the other is non-Markovian. \textcolor{blue}{$S_{M_1}$} and \textcolor{blue}{$S_{NM_1}$} are the two TLSs. \textcolor{blue}{$B_{M_1}$} is the Markovian bath, while \textcolor{blue}{$B_{NM_1}$} is non-Markovian one.
}
\label{fig:1}
\end{figure}

%\emph{\textbf{Multiparty GKSL-like equation for local Markovian and non-Markovian baths.--}}
\label{sec:2}
We consider $m+n$ subsystems locally coupled to $m$ Markovian and $n$ non-Markovian baths, respectively. 
%among which $m$ systems are weakly coupled to $m$ Markovian baths locally and the other $n$ are also locally interacting with $n$ non-Markovian baths. 
The situation is depicted in Fig.~\ref{fig:1}.
 The composite system consisting of $m+n$ subsystems will evolve under the combined influence of the local Markovian and non-Markovian baths. 
 %\textcolor{blue}{A detail discussion on the utility of studying this type of combined evolution is presented in the Appendix.} 
 Before considering the case of arbitrary $m$ and $n$, we deal with the case of two subsystems ($S_{M_1}$ and $S_{NM_1}$), locally coupled, respectively, to two heat baths, one of which is a Markovian bath ($B_{M_1}$) %\textcolor{red}{\sout{and which}}
that can be treated under the Born-Markov approximation,
%bosonic bath consisting of infinite number of harmonic oscillators 
while the other is a non-Markovian one ($B_{NM_1}$), whose frequency spectrum is discrete and goes beyond the Markovian regime.
%spin bath containing $N$ numbers of spin-$\frac{1}{2}$ particles centering the system-2 with equal distances. 
%\textcolor{red}{\sout{A schematic diagram of this}} 
This two-party two-bath setup is illustrated in the %\textcolor{red}{\sout{right panel of Fig.~$1$ of the Appendix}} 
grey box of Fig.~\ref{fig:1}.
The Hamiltonian of the composite setup is given by $H=H_s+H_B+H_I$, where $H_s$ describes the Hamiltonian of the composite system consisting of the two subsystems, $H_B$ stands for the combined local Hamiltonian of the two baths and $H_I=\sum_X H_{I_{X_1}}$ for $X \in \{M,NM\}$. Here $H_{I_{M_1}}$ represents the interaction between $S_{M_1}$ and $B_{M_1}$, and $H_{I_{NM_1}}$ presents the interaction between $S_{NM_1}$ and $B_{NM_1}$. Note that, the Hamiltonian $H_s$, describing the Hamiltonian of the composite system containing two subsystems, is a general Hamiltonian encompassing both the local and interacting part of the two subsystems. Precisely, this $H_s$ can be written as $H_s=H_{\text{loc}}+V$, where $H_{\text{loc}}$ is the local Hamiltonian of the two subsystems and $V$ represents the interaction between them. In the Schr\"odinger picture, let the density matrix of the composite two-party two-bath setup at time $t$ be represented by $\rho(t)$. It is useful to perform the calculation in the interaction picture~\cite{Petruccione}. 
%Now following the previous literature~\cite{Petruccione}, for the ease of calculation we move to the interaction picture. Hence, t
The von Neumann equation in that picture will be
\begin{equation}
\label{eq:liouville}
    \frac{d\rho_{\text{int}}(t)}{dt}=-\frac{i}{\hbar}\Big[H_{I_{\text{int}}}(t),\rho_{\text{int}}(t)\Big],
\end{equation}
where $H_{I_{\text{int}}}(t)=e^{\frac{i(H_s+H_B)t}{\hbar}}H_Ie^{-\frac{i(H_s+H_B)t}{\hbar}}$, $\rho_{\text{int}}(t)=e^{\frac{i(H_s+H_B)t}{\hbar}}\rho(t)e^{-\frac{i(H_s+H_B)t}{\hbar}}$, 
%\begin{eqnarray}
%    &&H_{I_{int}}(t)=e^{\frac{i(H_s+H_B)t}{\hbar}}H_Ie^{-\frac{i(H_s+H_B)t}{\hbar}},\nonumber\\
%    &&\rho_{int}(t)=e^{\frac{i(H_s+H_B)t}{\hbar}}\rho(t)e^{-\frac{i(H_s+H_B)t}{\hbar}},
%\end{eqnarray}
without assuming a commutativity relation of $H_I$ and $\rho(t)$ with $H_s$ and $H_B$. Here we assume that initially the systems are uncorrelated to the baths, and that the baths themselves are also in a product state, so that at time $t=0$, $\rho(0)=\rho_s(0)\otimes \rho_{B_{M_1}}(0) \otimes \rho_{B_{NM_1}}(0)$,
%\begin{equation}
%    \rho(0)=\rho_s(0)\otimes \rho_{B_1}(0) \otimes \rho_{B_2}(0),
%\end{equation}
where $B_{M_1}$ is initially in its stationary state, i.e., $[H_{B_{M_1}},\rho_{B_{M_1}}(0)]=0$, with $H_{B_{NM_1}}$ being the free Hamiltonian of the non-Markovian bath $B_{NM_1}$. %their canonical equilibrium states given by \textcolor{blue}{$\rho_{B_{X}}(0)=\frac{e^{-\beta_{X}H_{B_X}}}{\text{tr}(e^{-\beta_{X}H_{B_X}})}$ for $X \in \{M_1,NM_1\}$}.} 
The derivation of the GKSL-like equation for this two-party two-bath setup is given in Appendix~\ref{appen1}. %\textcolor{blue}{\sout{The interaction Hamiltonian in the Schr\"odinger picture can be decomposed in the form}~\cite{Petruccione,Rivas}, \stkout{$H_{I_j}=\sum_{k} A_{j_k} \otimes B_{j_k}$,}
As we mentioned earlier,  the bath $B_{M_1}$ is Markovian and therefore in the derivation, %\textcolor{blue}{\sout{further calculations}},
while talking about $B_{M_1}$, we have imposed %\textcolor{blue}{\sout{will impose}} 
the Born-Markov approximations, which tells that the coupling between the subsystem $S_{M_1}$ and $B_{M_1}$ is weak, so that the state of $B_{M_1}$ regains its initial state after every time step of interaction with $S_{M_1}$, and that any correlation created between $B_{M_1}$ and $S_{M_1}$ is also destroyed after the same time step. Moreover, $B_{M_1}$ will also be assumed to remain uncorrelated with $B_{NM_1}$ during the evolution. And also, the development of the reduced state of the system with respect to the bath \(B_{M_1}\), at each time, is assumed memoryless. %\textcolor{blue}{\sout{So, at time $t$, the state of the entire setup will take the form, $ \rho(t) \approx \rho_{sB_2}(t) \otimes \rho_{B_1}$,
The reduced system dynamics in the Schr\"odinger picture turns out to be
% \begin{eqnarray}
%     &&\frac{d\tilde{\rho}_{s}(t)}{dt}=\mathcal{L}\big(\tilde{\rho}_{s}(t)\big)\equiv\nonumber\\
%     &&-\frac{i}{\hbar}\Big[H_s,\tilde{\rho}_s(t)\Big]+\frac{1}{\hbar^2}\sum_{\omega}\sum_{k,l} \gamma_{k,l}(\omega) \Big(A_{1_l}(\omega)\tilde{\rho}_{s}(t)A_{1_k}^{\dagger}(\omega)\nonumber\\
 %  &&-\frac{1}{2}\big\{A_{1_k}^{\dagger}(\omega)A_{1_l}(\omega),\tilde{\rho}_{s}(t)\big\}\Big)-\frac{i}{\hbar}\text{tr}_{B_2}\Big[H_{I_{2}}(t),\rho_{sB_2}(t)\Big].\;\;\;\;\;\;\;
% \end{eqnarray}
%Here we have used that $\tilde{\rho}_s(t)=\text{tr}_{B_2}\big\{e^{\frac{iH_{B_2}t}{\hbar}}\rho_{sB_2}(t)e^{\frac{-iH_{B_2}t}{\hbar}}\big\}=\text{tr}_{B_2}(\rho_{sB_2})$. 
%\textcolor{red}{[(19) e use kora hoyechhe.]}
%Thus, we can write
\begin{eqnarray}
\label{eq:20}
    \frac{d\tilde{\rho}_{s}(t)}{dt}&=&\mathcal{L}\big[\tilde{\rho}_{s}(t)\big]\equiv -\frac{i}{\hbar}\Big[H_s,\tilde{\rho}_s(t)\Big]+\mathcal{D}_{M_1}\big[\tilde{\rho}_{s}(t)\big]\nonumber\\
    &&\phantom{ki gabo ami ki shuna}+\mathcal{D}_{NM_1}\big[\rho_{sB_2}(t)\big],
\end{eqnarray}
where
\begin{align}
    &\mathcal{D}_{M_1}\big[\tilde{\rho}_{s}(t)\big]=\frac{1}{\hbar^2}\sum_{\omega}\sum_{k,l} \gamma_{k,l}(\omega) \Big(A_{M_{1_l}}(\omega)\tilde{\rho}_{s}(t)A_{M_{1_k}}^{\dagger}(\omega)\nonumber\\
   &\phantom{amay akhash bollo}-\frac{1}{2}\big\{A_{M_{1_k}}^{\dagger}(\omega)A_{M_{1_l}}(\omega),\tilde{\rho}_{s}(t)\big\}\Big),\nonumber\\
   &\mathcal{D}_{NM_1}\big[\rho_{sB_{NM_{1}}}(t)\big]=-\frac{i}{\hbar}\text{tr}_{B_{NM_1}}\Big[H_{I_{NM_{1}}}(t),\rho_{sB_{NM_1}}(t)\Big].\;
   \label{eq:21}
\end{align}
%\textcolor{red}{
We denote $\tilde{\rho}_s(t)=\text{tr}_{B_{NM_1}}\Big[e^{\frac{iH_{B_{NM_1}}t}{\hbar}}\rho_{sB_{NM_1}}(t)e^{\frac{-iH_{B_{NM_1}}t}{\hbar}}\Big]\\=\text{tr}_{B_{NM_1}}\big[\rho_{sB_{NM_1}}(t)\big]$. Here $\omega$ is the transition energy, $A_{M_{1_k}}(\omega)$ are the Lindblad operators defined after Eq.~(\ref{eq:liouville_appen}), and $\gamma_{k,l}(\omega)$ are the transition rates defined after Eq.~(\ref{appen:A8}) in Appendix~\ref{appen1}.  Here we have neglected the possible effects of the Lamb-shift Hamiltonian in the dynamics~\cite{Ghoshal,Correa}. Effectively, $\mathcal{D}_{M_1}\big[\tilde{\rho}_{s}(t)\big]$ contains the dissipative part of the GKSL-like equation due to the influence of the Markovian bath and $\mathcal{D}_{NM_1}\big[\rho_{sB_{NM_1}}(t)\big]$ contains the non-Markovian contribution.
%$H_{LS}$, which contributes in the first term of Eq.~(\ref{eq:20}) as $-\frac{i}{\hbar}\Big[H_s+H_{LS},\tilde{\rho}_s(t)\Big]$, as we resides in the weak coupling limit $\gamma_{k,l}(\omega) \ll$ the parameters of $H_s$, for the Markovian evolution, generally satisfied in the optical quantum regime~\cite{Ghoshal,Correa}.} 
Interestingly, although the bath $B_{M_1}$ is treated as Markovian, the dissipator $\mathcal{D}_{M_1}\big[\tilde{\rho}_s(t)\big]$ of Eq.~(\ref{eq:21}) reveals a crucial distinction between our approach and the traditional Markovian cases~\cite{Petruccione,Rivas}. In the Markovian scenario, the dissipative term would involve the system's state uncorrelated with any environment. However, in our formulation, this term incorporates $\tilde{\rho}_s(t)$, which represents the system's state correlated with the non-Markovian bath. The correlation between the system and non-Markovian baths implies that the composite state of the system and the non-Markovian bath at time $t$ cannot be expressed as a product state, unlike in the case of Markovian baths. Therefore, a crucial distinction arises due to the presence of the non-Markovian effect of $B_{NM_1}$, which persists within the contribution of $B_{M_1}$. %\textcolor{red}{\sout{As the bath \textcolor{blue}{$B_{M_1}$} is Markovian, we can expect that the effect coming from the bath in the evolution of the system should be completely Markovian. However, we see from Eq.~(\ref{eq:21}) that the term capturing the contribution of the Markovian bath contains $\tilde{\rho}_s(t)$, the system's state correlated with the non-Markovian bath, instead of the system's state, uncorrelated with any environment, as it occurs in the Markovian approach. This is a major difference between the Markovian case~\cite{Petruccione,Rivas} and the situation %\textcolor{blue}{considered in this letter.} which we have referred to as mixed set of local environments. Although we have considered \textcolor{blue}{$B_{M_1}$} as a Markovian bath, the non-Markovian effect of \textcolor{blue}{$B_{NM_1}$} remains in the contribution of \textcolor{blue}{$B_{M_1}$}.}}
Therefore, both $\mathcal{D}_{M_1}\big[\tilde{\rho}_s(t)\big]$ and $\mathcal{D}_{NM_1}\big[\rho_{sB_{NM_1}}(t)\big]$ serve as non-Markovian dissipators for this combined evolution. 
%\textcolor{red}{\sout{$\mathcal{D}_M(\tilde{\rho}_s(t))$ can be named as a Markovian-like dissipator, whereas \textcolor{blue}{$\mathcal{D}_{NM}(\rho_{sB_{NM_1}}(t))$} can be called as a non-Markovian dissipator.}}
\par 
%Now we need to discuss about the thermodynamic quantities for the Markovian and non-Markovian combined evolution.\par
%We now move over to the case of an arbitrary number of subsystems, as schematically depicted in the left panel of Fig.~\ref{fig:1}.
The two-party dynamical equation given in Eq.~(\ref{eq:20}) can be extended to the situation where $m+n$ subsystems are connected to $m$ Markovian and $n$ non-Markovian baths locally (see Fig.~\ref{fig:1}). For that general case, the dynamical equation of the system takes the form,
\begin{eqnarray}
\label{eq:22}
    &&\mathcal{L}\big[\tilde{\rho}_{s}(t)\big]=-\frac{i}{\hbar}\Big[H_s,\tilde{\rho}_s(t)\Big]+\sum_{j=1}^m\mathcal{D}_{M_j}\big[\tilde{\rho}_{s}(t)\big]\nonumber\\
    &&\phantom{mone robe ki na ro}+\sum_{j=1}^n\mathcal{D}_{NM_j}\big[\rho_{sB_{NM_{1\dots n}}}(t)\big].
\end{eqnarray}
Here $\tilde{\rho}_{s}(t)=\text{tr}_{B_{NM_{1\dots n}}}\big(\rho_{sB_{NM,1\dots n}}(t)\big)$. The $\mathcal{D}_{M_j}\big[\tilde{\rho}_{s}(t)\big]$
%\textcolor{red}{\sout{'s have the same form as $\mathcal{D}_M(\tilde{\rho}_{s}(t))$, presenting}} 
presents the contribution of the $j^{\text{th}}$ Markovian bath. Similarly, $\mathcal{D}_{NM_j}\big[\rho_{sB_{NM_{1\dots n}}}(t)\big]$ represents the contribution of the $j^{\text{th}}$ non-Markovian bath.
%like $\mathcal{D}_{NM}(\rho_{sB_2}(t))$, 
%only the state becomes correlated with all non-Markovian baths.
With the increase in the number of non-Markovian baths, the system in general will tend to become more and more correlated with the non-Markovian baths, but the effect of the Markovian baths will also in general become significantly altered in comparison to the situation where non-Markovian environments are absent. This in turn may affect the general properties and inter-relations between thermodynamic quantities that are typically considered in either Markovian or non-Markovian situations before.  
%\textcolor{red}{\sout{non-mixed sets of local environs.}} 
Below, we study the response of the heat current and EPR, and their inter-relation via the Spohn's theorem, %\textcolor{red}{\sout{to the incorporation of mixed sets of local baths.}} 
for the combination of local Markovian and non-Markovian dynamics.\par
%We can realise that, with the increase of the number of non-Markovian bath the state of the system becomes more correlated with the baths. This correlation, arises due to the presence of non-Markovian baths along with the Markovian baths in a combined system bath setup, may effect the general definitions and realisations of some thermodynamic quantities. Hence the study of thermodynamic quantities like heat current, EPR is necessary in combined local Marvian and non-Marvian evolution scenario. In the next part of the paper we discuss the effect of the aforementioned situation on some thermodynamic quantities and obtain a modification of Spohn's theorem which can be considered as an important ingredient in quantum thermodynamics.

\section{Thermodynamic quantities and their properties for combination of local Markovian and non-Markovian %\textcolor{red}{\sout{mixed}} 
environs}
%\emph{\textbf{Thermodynamic quantities and their properties for combination of local Markovian and non-Markovian %\textcolor{red}{\sout{mixed}} 
%environs.--}}
\label{sec:3}
Heat current and EPR are two fundamental thermodynamic properties of a system which provide information about heat flow from the system to its environment or vice versa, and further aspects of equilibrium and non-equilibrium physics of the system. 
%If the system resides in its equilibrium steady state after an evolution in presence of a Markovian environment, EPR takes the value 0 and for the situation where the system reaches a non-equilibrium steady state, EPR is always $>0$. 
It is known that for a non-Markovian evolution, EPR can take a negative value~\cite{Bylicka,Marcantoni,Bhattacharya,Popovic,Strasberg1,Naze,Bonanca,Ghoshal} and as a consequence, it can be treated as a witness of non-Markovianity. The definitions of the thermodynamic quantities can strongly depend on the character of the environments under which the system is being evolved.
%For example, under local non-Markovian evolutions, the general definitions of heat current and EPR differ from that in the ideal Markovian situation~\cite{Ghoshal}.\par
%Our situation in this paper is a bit different from the case considered in~\cite{Ghoshal}.
In general, entropy flux or heat current can be defined as the amount of entropy exchanged per unit time between the open system and its environment~\cite{Petruccione}.
%, and can be defined as
%. because they are equivalent. In this paper we have considered this entropy flux as heat current. 
%Here we assume 
%For positive values of the heat current, heat flows from the system to the environment, and for negative values of the same, the direction of heat flow will be the opposite. 
Entropy flux for the composite two-party system depicted in the %\textcolor{red}{\sout{right panel}}
grey box of Fig.~\ref{fig:1} can be defined as $J_{\{M_1,NM_1\}}=\frac{d}{dt}\Big|_{\text{diss}}E$,
%\begin{equation}
 %   J_{12}=\frac{d}{dt}\Big|_{diss}E,
%\end{equation}
which indicates the change in internal energy resulting from the dissipative effects. Here $E=\text{tr}\big[{H_s\tilde{\rho}_{s}(t)}\big]$. So,
\begin{eqnarray}
\label{eq:24}
    J_{\{M_1,NM_1\}}&=&\text{tr}\Big[H_s \Big(\mathcal{D}_{M_1}\big[\tilde{\rho}_s(t)\big]+\mathcal{D}_{NM_1}\big[\rho_{sB_{NM_1}}(t)\big]\Big)\Big]\nonumber\\
    &&\phantom{ami pothbhola ek po}=\text{tr}\Big[H_s\mathcal{L}\big[\tilde{\rho}_{s}(t)\big]\Big].
\end{eqnarray}
We define the local heat currents of each subsystem as $J_{M_1}=\text{tr}\big[H_s \mathcal{D}_{M_1}\big[\tilde{\rho}_s(t)\big]\big]$ and $J_{NM_1}=\text{tr}\big[H_s \mathcal{D}_{NM_1}\big[\rho_{sB_{NM_1}}(t)\big]\big]$.
%\begin{eqnarray}
 %   J_1&=&\text{tr}\big(H_s \mathcal{D}_M(\tilde{\rho}_s(t))\big),\nonumber\\
  %  J_2&=&\text{tr}\big(H_s \mathcal{D}_{NM}(\rho_{sB_2}(t))\big).
%\end{eqnarray}
The definition of $J_{M_1}$ is quite similar to that in the Markovian approach, but the effect of non-Markovianity resides in the state $\tilde{\rho}_s(t)$, as the system's state is correlated with the non-Markovian bath. The formulation of  $J_{NM_1}$ is inherently non-Markovian.\par 
EPR is a thermodynamic quantity of a system, which is defined as a source term in the balance equation involving the rate of change of entropy with time and heat current~\cite{Spohn,Lebowitz,Groot}. For a two-party two-bath composite setup, this balance equation can be considered as
\begin{equation}
\sigma_{\{M_1,NM_1\}}=\frac{dS(\tilde{\rho}_s(t))}{dt}-\frac{1}{T_{M_1}} J_{M_1}-\frac{1}{T_{NM_1}} J_{NM_1},
    \label{eq:26}
\end{equation}
where $S(\cdot)$ is the von Neumann entropy of its argument and defined as $S(\rho)=-k_B\text{tr}\big[\rho \ln(\rho)\big]=-k_B\sum_i \lambda_i \ln (\lambda_i)$, with $\lambda_i$'s being the eigenvalues of the density matrix $\rho$. $T_{M_1}$ and $T_{NM_1}$ are the temperatures of the Markovian and non-Markovian baths, respectively. $k_B$ is the Boltzmann constant. In this formulation, the balance equation is the definition of the EPR, denoted by $\sigma_{\{M_1,NM_1\}}$. 
%\textcolor{red}{\sout{, with the suffixes indicating the systems involved. In Markovian dynamics of a single system, EPR turns out to be the negative derivative of the relative entropy distance from the canonical equilibrium state of the system~\cite{Spohn,Lebowitz}, but for non-Markovian cases, this interpretation may not apply~\cite{Ghoshal}.}}
\par
We now move over to the case of $m+n$ subsystems (see 
%\textcolor{red}{\sout{the left panel of}} 
Fig.~\ref{fig:1}). For $m+n$ parties, the global heat current and the global EPR take the following forms:
\begin{eqnarray}
%&&\textcolor{red}{\stkout{J_{\{M_{1 \cdots m},NM_{1 \cdots n}\}}=\sum_{j=1}^m\text{tr}\Big(H_s (\mathcal{D}_{M_j}(\tilde{\rho}_s(t))}}\nonumber\\
%&&\phantom{ami keboli swapono}\textcolor{red}{\stkout{+\sum_{j=1}^n\mathcal{D}_{NM_{j}}(\rho_{sB_{NM_{1 \cdots n}}}(t))\big)\Big),}}\nonumber\\
&&J_{\{M_{1 \cdots m},NM_{1 \cdots n}\}}=\sum_{j=1}^m J_{M_j}+\sum_{j=1}^n J_{NM_j},\nonumber\\
&&\sigma_{\{M_{1 \cdots m},NM_{1 \cdots n}\}}=\frac{dS(\tilde{\rho}_s(t))}{dt}-\sum_{j=1}^{m}\frac{1}{T_{M_j}} J_{M_j}\nonumber\\ 
&&\phantom{aj aloy aloy akash dekho aka}-\sum_{j=1}^{n}\frac{1}{T_{NM_j}} J_{NM_j}.
\end{eqnarray}
%\textcolor{red}{$J_{total}=\sum_{k=1}^m J_{M,k}+\sum_{l=1}^n J_{NM,l}$} 
Here the local heat currents, $J_{M_j}=\text{tr}\Big[H_s \mathcal{D}_{M_j}\big[\tilde{\rho}_s(t)\big]\Big]$, presents the heat current flowing to or from the $j^{\text{th}}$ Markovian bath 
%is presented by the $j^{\text{th}}$ contribution of the first term in \textcolor{blue}{$J_{\{M_{1 \cdots m},NM_{1 \cdots n}\}}$} 
for $j=1,\cdots,m$, and the 
%same of the 
second term, $J_{NM_j}=\text{tr}\Big[H_s\mathcal{D}_{NM_{j}}\big[\rho_{sB_{NM_{1 \cdots n}}}(t)\big]\Big]$, signifies the local heat currents
%, \textcolor{blue}{$J_{NM_j}=\text{tr}\big(H_s\mathcal{D}_{NM_{j}}(\rho_{sB_{NM_{1 \cdots n}}}(t))\big)$}, 
flowing towards or outwards from the $j^{\text{th}}$ non-Markovian bath for $j=1, \cdots, n$.\par
%and the second terms of $J_{1 \cdots (m+n)}$. The first term represents the local heat currents comes from the $i^{\text{th}}$ Markovian bath and the second one present the same for the $i^{th}$ non-Markovian bath.
%$\sigma_{12}$ is the entropy production rate (EPR) of the composite two-qubit system.
%\begin{equation}
 %   \sigma_{12}=-\frac{d}{dt}S\big(\tilde{\rho}_s(t)||\rho_{th}\big)-k_B\text{tr}\big(\tilde{\rho}_s\mathcal{L}(ln(\rho_{th}))\big)
%\end{equation}
\noindent
\section{Modification to Spohn's theorem}
%\emph{\textbf{Modification to Spohn's theorem.--}}
\label{sec:4}
For the evolution of a system under a Markovian reservoir, which is initially kept in its canonical equilibrium state, Spohn's theorem assures the positivity of EPR,  %\textcolor{red}{\sout{for a Markovian process.}} 
as under this circumstance, the canonical equilibrium state of the system is a stationary state with respect to the Markovian dynamics. 
%This is a well known result and first derived by H. Spohn in~\cite{Spohn,Lebowitz}. 
For non-Markovian evolutions, the positivity of the same quantity may hold, but is not guaranteed even if the bath is initially in its thermal state. 
%It is already known that for some non-Markovian processes EPR may take negative values~\cite{Bylicka,Marcantoni,Bhattacharya,Popovic,Strasberg1,Naze,Bonanca,Ghoshal}.
%but for a local or global non-Markovian process EPR can take negative values. 
So, there may exist a modified form of Spohn's theorem for non-Markovian evolutions which can describe the thermodynamics of non-Markovian scenarios. Our aim is to obtain a thermodynamic condition similar to that in the Spohn's theorem, which can describe a multiparty situation with a combination of local Markovian and non-Markovian %\textcolor{red}{\sout{mixed set of local}} 
environs.
\par
%Here we try to derive a modified form of the theorem for local Markovian and non-Markovian combined evolution.
%First we consider a system-bath setup which is not purely Markovian. 
For the ease of notation and calculations, we take the simple two-party two-bath situation depicted in the gray box 
%\textcolor{red}{\sout{right panel}}
of Fig.~\ref{fig:1}. The dynamical equation of the system for this setup is given in Eq.~(\ref{eq:20}). We %\textcolor{red}{\sout{separate the Markovian-like part and the non-Markovian one from the GKSL equation (Eq.~(\ref{eq:20}))  and}} 
define the partial superoperators~\cite{Hewgill},
\begin{eqnarray}
&&\mathcal{L}_{M_1}\big[\tilde{\rho}_s(t)\big]=-\frac{ip}{\hbar}[H_s,\tilde{\rho}_s(t)]+\mathcal{D}_{M_1}\big[\tilde{\rho}_s(t)\big], \nonumber\\
    &&\mathcal{L}_{NM_1}\big[\rho_{sB_{NM_1}}(t)\big]=-\frac{i(1-p)}{\hbar}[H_s,\tilde{\rho}_s(t)]\nonumber \\
    &&\phantom{borne gondhe chhonde gitite}+\mathcal{D}_{NM_1}\big[\rho_{sB_{NM_1}}(t)\big],
    %\;\;\;\;\;\;\;\;
\end{eqnarray}
where 
%\textcolor{red}{\sout{the total Lindblad operator is}} 
$\mathcal{L}\big[\tilde{\rho}_s(t)\big]=\mathcal{L}_{M_1}\big[\tilde{\rho}_s(t)\big]+\mathcal{L}_{NM_1}\big[\rho_{sB_{NM_1}}(t)\big]$ and $p$ is a weight factor. The two parts $\mathcal{L}_{M_1}\big[\tilde{\rho}_s(t)\big]$ and $\mathcal{L}_{NM_1}\big[\rho_{sB_{NM_1}}(t)\big]$ act as GKSL-like equation operators individually, with $\text{tr}\big[\mathcal{L}_{M_1}\big[\tilde{\rho}_s(t)\big]\big]=\text{tr}\big[\mathcal{L}_{NM_1}\big[\rho_{sB_{NM_1}}(t)\big]\big]=0$. Hence, the local heat currents can be described in terms of the local superoperators as $J_{M_1}=\text{tr}\big[H_s \mathcal{L}_{M_1}\big[\tilde{\rho}_s(t)\big]\big]$, $J_{NM_1}=\text{tr}\big[H_s \mathcal{L}_{NM_1}\big[\rho_{sB_{NM_1}}(t)\big]\big]$.
%\begin{eqnarray}
 %   J_1&=&\text{tr}\big(H_s \mathcal{L}_M(\tilde{\rho}_s(t))\big),\nonumber\\
  %  J_2&=&\text{tr}\big(H_s \mathcal{L}_{NM}(\rho_{sB_2}(t))\big).
%\end{eqnarray}
Now we introduce two ``local" canonical equilibrium states of the composite system at temperatures $T_{M_1}$ and $T_{NM_1}$~\cite{Hewgill} having the form, $\tilde{\rho}_{\text{th}_{X_1}}=\frac{e^{-\beta_{X_1} H_s}}{Z_{X_1}}$,
%which are not exactly the cannonical equilibrium state for the composite system, but can be the cannonical equilibrium state 
%So, The cannonical equilibrium states at temperature $T_1$ and $T_2$ are respectively,
%\textcolor{blue}{\begin{equation}
%\label{eq:can}
    %\tilde{\rho}_{th_X}=\frac{e^{-\beta_X H_s}}{Z_X},
%\end{equation}}
%\textcolor{red}{\sout{for $j=1$ and 2}} 
for $X \in \{M, NM\}$. Note that $H_s$ is a Hamiltonian of two parties. $Z_{X_1}$'s stand for the corresponding partition functions and defined as $Z_{X_1}=\text{tr}(e^{-\beta_{X_1} H_s})$. Here $\beta_{X_1}=\frac{1}{k_BT_{X_1}}$. Thus, using $\tilde{\rho}_{th_{X_1}}$ %\textcolor{red}{\sout{Eq.~(\ref{eq:can})}} 
and the partial superoperators, we get
\begin{eqnarray}
    &&\sigma_{\{M_1,NM_1\}}=
    %-k_B\Big[\text{tr}\big(\mathcal{L}(\tilde{\rho}_s(t))\ln(\tilde{\rho}_s(t))\big)-\text{tr}\big(\mathcal{L}_M(\tilde{\rho}_s(t))\ln(\tilde{\rho}_{th_1})\big)\nonumber\\
    %&&\phantom{kotobaro bhebechh}-\text{tr}\big(\mathcal{L}_{NM}(\rho_{sB_2}(t))\ln(\tilde{\rho}_{th_2})\big)]\Big]\nonumber\\
    -\frac{d}{dt}\Big|_M S\big(\tilde{\rho}_s(t)||\tilde{\rho}_{{\text{th}}_{M_1}}\big)\nonumber\\
    &&-k_B\text{tr}\Big[\mathcal{L}_{NM_1}\big[\rho_{sB_{NM_1}}(t)\big]\big(\ln(\tilde{\rho}_s(t))-\ln(\tilde{\rho}_{\text{th}_{NM_1}})\big)\Big],\;\;\;\;\;\;\;
    \label{eq:sigma}
\end{eqnarray}
where the relative entropy distance, $S(\rho||\sigma)=k_B\text{tr}(\rho\ln \rho-\rho\ln\sigma)$, is used to quantify the ``distance" between the evolved state and the local canonical equilibrium state at temperature $T_{M_1}$, and $\frac{d}{dt}\Big|_M S\big(\tilde{\rho}_s(t)||\tilde{\rho}_{{\text{th}}_{M_1}}\big)=k_B\text{tr}\big\{\mathcal{L}_{M_1}\big[\tilde{\rho}_s(t)\big]\big(\ln(\tilde{\rho}_s(t))-\ln(\tilde{\rho}_{\text{th}_{M_1}})\big)\big\}$.
%\textcolor{red}{\sout{, which is a stationary state for the local superoperator $\mathcal{L}_M(\cdot)$, i.e., $\mathcal{L}_M(\tilde{\rho}_{th_{M_1}})=0$. As we can see, the first term in~(\ref{eq:sigma}) is a contribution of the Markovian bath, so that $\frac{d}{dt}\Big|_M$ denotes a Markovian-like contribution, $\mathcal{L}_M(\cdot)$,  to the total GKSL equation.}}
%, which is $\mathcal{L}_M(\cdot)$.
 In the Markovian limit of the setup under consideration, i.e., when both the baths are Markovian, the first term of Eq.~(\ref{eq:sigma}) will be duplicated for the other bath, and the second term will be non-existent.
%For a pure Markovian evolution only the first term appears and the second term vanishes~\cite{Petruccione,Spohn,Lebowitz}. 
Hence, Eq.~(\ref{eq:sigma}) can be presented as a general expression of EPR for a two-party system evolving under a combination of local Markovian and non-Markovian dynamics.%\textcolor{red}{\sout{mixed set of environments.}}
\par 
%This kind of modification in the definition of EPR for a different non-Markovian scenario was obtained in~\cite{Ghoshal}. There the authors have considered a local non-Markovian dynamics of a global Markovian evolution.
%This is to be contrasted with the modified EPR obtained in~\cite{Ghoshal} for the local non-Markovian evolution of a globally Markovian dynamics.\par
We now try to establish the Spohn's theorem with the altered definition of EPR. From Eq.~(\ref{eq:sigma}), we can write
\begin{eqnarray}
\label{eq:NM1}
   &&\sigma_{\{M_1,NM_1\}}\nonumber\\
   &&+k_B\text{tr}\big\{\mathcal{L}_{NM_1}\big[\rho_{sB_{NM_1}}(t)\big]\big(\ln(\tilde{\rho}_{s}(t))-\ln(\tilde{\rho}_{\text{th}_{NM_1}})\big)\big\}\nonumber\\
   &&\phantom{at}=-k_B\text{tr}\big\{\mathcal{L}_{M_1}\big[\tilde{\rho}_s(t)\big]\big(\ln(\tilde{\rho}_s(t))-\ln(\tilde{\rho}_{\text{th}_{M_1}})\big)\big\}.\;\;\;
\end{eqnarray}
If the initial state of the Markovian bath is the canonical equilibrium state, then the state $\tilde{\rho}_{\text{th}_{M_1}}$ will be the stationary state with respect to $\mathcal{L}_{M_1}[\cdot]$, i.e., $\mathcal{L}_{M_1}\big[\tilde{\rho}_{\text{th}_{M_1}}\big]=0$. Spohn's inequality~\cite{Spohn} tells us that for any superoperator of Lindblad form, say $\mathcal{L}_{M_1}[\cdot]$, with a stationary state, say $\tilde{\rho}_{\text{th}_{M_1}}$, the right-hand side (RHS) of Eq.~(\ref{eq:NM1}) is always $\ge 0$. On the contrary, when one must go beyond the Markovian approximations while considering the dynamics of a system, the existence of a steady state is not guaranteed. Moreover, in the second term of the left-hand side (LHS) of~(\ref{eq:NM1}), $\tilde{\rho}_{\text{th}_{NM_1}}$ will in general not be the steady state corresponding to $\mathcal{L}_{NM_1}[\cdot]$ irrespective of the initial state of the non-Markovian bath. So, we cannot infer the sign of that term, as can, e.g., be seen for the case involving four qubits undergoing a combined local Markovian and non-Markovian evolution presented in Appendix~\ref{appen4}, where we show that this term can take both positive and negative values. Thus, for the case of a two-party system with the subsystems being attached separately to two baths, one of which is Markovian and the other not, we get the altered form of Spohn's theorem as
\begin{align}
  &\sigma_{\{M_1,NM_1\}}
  \nonumber\\
  &+k_B\text{tr}\big\{\mathcal{L}_{NM_1}\big[\rho_{sB_{NM_1}}(t)\big]\big(\ln(\tilde{\rho}_s(t))-\ln(\tilde{\rho}_{\text{th}_{NM_1}})\big)\big\} 
  \ge 0, \nonumber\\
  \label{eq:33}
\end{align}
%with $M_{NM}^1(\rho_{sB_{NM_1}}(t))=k_B\text{tr}\big\{\mathcal{D}_{NM_1}(\rho_{sB_{NM_1}}(t))\big(\ln(\tilde{\rho}_s(t))-\ln(\tilde{\rho}_{th_{NM_1}})\big)\big\}$, 
provided that the Markovian bath is initially in its canonical equilibrium state. The second term in the LHS of the inequality is an extra term that has got appended due to the presence of the non-Markovian bath in the 
%\textcolor{red}{\sout{mixed}} 
set of local environments. 
%which makes a difference between the pure Markovian and the local Markovian and non-Markovian combined evolution.
%\textcolor{red}{\sout{The existence of a steady state for a Markovian evolution of Lindblad form is guaranteed. The canonical equilibrium state of the system is a steady state for the dynamics of the system under a Markovian environment. When one must go beyond the Markovian approximations while considering the dynamics of a system, the existence of a steady state is not guaranteed.  Moreover, in the second term of the L.H.S. of~(\ref{eq:33}), \textcolor{blue}{$\tilde{\rho}_{th_{NM_1}}$} will in general not be the steady state corresponding to \textcolor{blue}{$\mathcal{L}_{NM}(\rho_{sB_{NM_1}}(t))$}. So, we cannot infer the sign of that term and hence}} 
Hence, a modified version of Spohn's theorem arises, which states that not the EPR, but the conjunction of EPR and $M_{NM}^1\big[\rho_{sB_{NM_1}}(t)\big]=k_B\text{tr}\big\{\mathcal{D}_{NM_1}\big[\rho_{sB_{NM_1}}(t)\big]\big(\ln(\tilde{\rho}_s(t))-\ln(\tilde{\rho}_{\text{th}_{NM_1}})\big)\big\}$ is assured to be positive for a combination of local Markovian and non-Markovian %\textcolor{red}{\sout{mixed set of local}} 
environments. The $\mathcal{L}_{NM_1}\big[\rho_{sB_{NM_1}}(t)\big]$ in~(\ref{eq:33}) can be replaced by $\mathcal{D}_{NM_1}\big[\rho_{sB_{NM_1}}(t)\big]$, as the first term of the local superoperator $\mathcal{L}_{NM_1}\big[\rho_{sB_{NM_1}}(t)\big]$ has no contribution in $M_{NM}^1\big[\rho_{sB_{NM_1}}(t)\big]$. The presence of $M_{NM}^1\big[\rho_{sB_{NM_1}}(t)\big]$ in the inequality~(\ref{eq:33}), therefore, indicates a deviation from the Markovian regime. %\textcolor{red}{\sout{and can be regarded as a witness of non-Markovianity. 
%\begin{eqnarray}
%M_{NM}&=&k_B\text{tr}\big\{\mathcal{D}_{NM}(\rho_{sB_2}(t))\big(\ln(\tilde{\rho}_s(t))-\ln(\tilde{\rho}_{th_2})\big)\big\}.\;\;\;\;
%&=&\text{tr}\big\{\mathcal{D}_{NM}(\rho_{sB_2}(t))\big(\ln(\tilde{\rho}_s(t))-\ln(\tilde{\rho}_{th_2})\big)\big\}
%\end{eqnarray}
%A \textcolor{red}{\sout{nonzero}} \textcolor{blue}{positive} value of $M_{NM}^1$ signals non-Markovianity in the dynamics.}}
\par
%\textcolor{red}{\sout{The altered form of Spohn's theorem illustrated in~(\ref{eq:33}) is obtained for a two-qubit two-bath setup.}} 
For a general $(m+n)$ subsystems, schematically depicted in %\textcolor{red}{\sout{the left panel of}} 
Fig.~\ref{fig:1}, the 
%\textcolor{red}{\sout{total}} 
GKSL-like equation takes the form, $\frac{d}{dt}\tilde{\rho}_s(t)=\mathcal{L}\big[\tilde{\rho}_s(t)\big]\equiv\sum_{j=1}^m\mathcal{L}_{M_j}\big[\tilde{\rho}_s(t)\big]+\sum_{j=1}^n\mathcal{L}_{NM_j}\big[\rho_{sB_{NM_{1 \cdots n}}}(t)\big]$. The modified Spohn's theorem %\textcolor{red}{\sout{witness of non-Markovianity}} 
in this general case of $m+n$ parties turns out to be
%\textcolor{red}{\begin{eqnarray}
%\label{NM_n}
%&&\stkout{M_{NM}^n=k_B \sum_{j=1}^n \text{tr}\big\{\mathcal{D}_{NM_j}(\rho_{sB_{NM_{1\cdots n}}}(t))}\nonumber\\
%&&\phantom{aji jhoro jhoro muk}\stkout{\big(\ln(\tilde{\rho}_s(t))-\ln(\tilde{\rho}_{th_{NM_j}})\big)\big\}}
%&&+\text{tr}\big\{\mathcal{D}_{NM_2}(\rho_{sB_{NM,1\dots n}}(t))\big(\ln(\tilde{\rho}_s(t))-\ln(\tilde{\rho}_{th_{NM_2}})\big)\big\}+\dots\nonumber\\
%&&+\text{tr}\big\{\mathcal{D}_{NM_n}(\rho_{sB_{NM,1\dots n}}(t))\big(\ln(\tilde{\rho}_s(t))-\ln(\tilde{\rho}_{th_{NM_n}})\big)\big\}.\;\;\;\;\;\;
%\end{eqnarray}}
%\textcolor{red}{\sout{and the modified Spohn's theorem has the form,}}
\begin{equation}
\sigma_{\{M_{1 \cdots m},NM_{1 \cdots n}\}}+\sum_{j=1}^nM_{NM}^j\big[\rho_{sB_{NM_{1\cdots n}}}(t)\big] \ge 0,
\label{NM_n1}
\end{equation}
%where $M_{NM}^n(\rho_{sB_{NM_{1\cdots n}}}(t))=k_B \sum_{j=1}^n \text{tr}\big\{\mathcal{D}_{NM_j}(\rho_{sB_{NM_{1\cdots n}}}(t))
%\big(\ln(\tilde{\rho}_s(t))-\ln(\tilde{\rho}_{th_{NM_j}})\big)\big\}$. 
%with $M_{NM}^j(\rho_{sB_{NM_{1\cdots n}}}(t))=\text{tr}\big\{\mathcal{D}_{NM_j}(\rho_{sB_{NM_{1\cdots n}}}(t))\big(\ln(\tilde{\rho}_s(t))-\ln(\tilde{\rho}_{th_{NM_j}})\big)\big\}$
%As we previously mentioned,  $n$ is the number of non-Markovian baths here. In other words, $n$ can be thought of as the number of partial superoperators associated with a, for which the local canonical equilibrium states $\tilde{\rho}_{th_{NM_j}}$ is not the stationary state when the environments are kept in their canonical equilibrium state initially. Therefore, 
for all Markovian baths being kept in their canonical equilibrium states at $t=0$. Here, the number of non-Markovian baths, $n$, can be interpreted as the count of partial superoperators for which the corresponding states $\tilde{\rho}_{\text{th}_{NM_j}}$ are not the stationary states. 
Conversely, if all the baths are Markovian, then $\tilde{\rho}_{\text{th}_{M_j}}$ will represent the stationary states of their respective partial superoperators for all $j$, resulting in $n=0$. Hence, 
%i.e., for $n=0$, 
Eq.~(\ref{NM_n1}) reverts to, $\sigma_{\{M_{1 \cdots m},NM_{1 \cdots n}\}}\ge 0$, which is the original form of the Spohn's theorem. 
%So, the altered form of the Spohn's theorem given in Eq.~(\ref{NM_n1}) reduces to its original form for $n=0$. 
%An important observation to note is that this modification of Spohn's theorem is valid only for the specific case of combined local evolution under both Markovian and non-Markovian environments. If we relax the restrictions imposed on the Markovian baths and consider all the environments as non-Markovian (i.e., $m=0$), then this type of modification of Spohn's theorem, as presented in Eq.~(\ref{NM_n1}), is no longer valid. 
On the other hand, in case all baths are non-Markovian, we obtain an altered form of the Spohn’s theorem that follows directly from the balance equation and the concept of EPR. Appendix~\ref{appen2} contains a detailed discussion on this matter.

In the context of inequality~(\ref{NM_n1}), we can introduce a witness for detecting non-Markovian behavior, as well as a measure for the same. Let us consider a situation where we have $q$ parties, each connected to $q$ environments locally, and the initial states of these environments are the respective canonical equilibrium states. Now, we evaluate the quantity $M^{k}\big[\rho^{\prime}(t)\big]=k_B\text{tr}\big\{\mathcal{D}_k\big[\rho^{\prime}(t)\big]\big(\ln(\tilde{\rho}_s(t))-\ln(\tilde{\rho}_{\text{th}_{k}})\big)\big\}$ associated with the dissipators $\mathcal{D}_k\big[\rho^{\prime}(t)\big]$ coming from the $k^{\text{th}}$ environment where $k$ runs from $1$ to $q$. The form of $\rho^{\prime}(t)$ depends on whether the environment associated with the dissipator is Markovian or non-Markovian. If the environment is Markovian, then $\rho^{\prime}(t)$ equals $\tilde{\rho}_s(t)=\text{tr}_{B_{1\cdots q}}\big[\rho_{sB_{1 \cdots q}}(t)\big]$. If the environment is non-Markovian, then $\rho^{\prime}(t)$ equals $\rho_{sB_{1\cdots q}}(t)$, representing the composite state of the systems and the baths. This construction of $\rho^{\prime}(t)$ is possible as we can replace the trace taken over the non-Markovian baths, denoted as $\text{tr}_{B_{NM_{1\cdots n}}}$ in Eq.~(\ref{eq:22}), with the trace taken over all the baths, denoted as $\text{tr}_{B_{1\cdots q}}$ while constructing the dissipators, because tracing out the Markovian baths has no impact on the dissipators, as they are product states with the remaining part of the system-bath setup. Therefore, in this $q$-party scenario, we can use $\rho_{sB_{1\cdots q}}$ instead of $\rho_{sB_{1\cdots n}}$. With the above definitions, we can now define a quantity $\overline{M}\big[t,\rho_s(0)\big]=\sum_{k=1}^{q}\max\{0,M^{k}\big[\rho^{\prime}(t)\big]\}$. If all the baths are Markovian, $\overline{M}\big[t,\rho_s(0)\big]$ will be zero. However, if at least one bath is non-Markovian, $\overline{M}\big[t,\rho_s(0)\big]$ can take values greater than zero. Hence, this quantity $\overline{M}\big[t,\rho_s(0)\big]$ serves as a witness of non-Markovianity as it detects the deviation of the altered Spohn's theorem for the combined local Markovian and non-Markovian dynamics from the original version of Spohn's theorem. Note that it is crucial to start with environments initially in their canonical equilibrium states. If the environments do not begin in these states, then $\overline{M}\big[t,\rho_s(0)\big]$ can yield positive values even for Markovian environments.
%denotes the  aswhen all the environments are kept in their thermal equilibrium state initially. 
%A \textcolor{blue}{positive} value of $\overline{M}^n_{NM}$ signals non-Markovianity in the dynamics.}
We can therefore define %\textcolor{blue}{ $M^n_{NM}$ as a witness} 
a quantifier of non-Markovianity
as
%\textcolor{red}{\begin{eqnarray}
%&&\stkout{\overline{M}_{NM}^n=k_B \sum_{j=1}^n\Big|\text{tr}\big\{\mathcal{D}_{NM_j}(\rho_{sB_{NM_{1\dots n}}}(t))}\nonumber\\
%&&\stkout{\phantom{modhyorate dakle}\big(\ln(\tilde{\rho}_s(t))-\ln(\tilde{\rho}_{th_{NM_j}})\big)\big\}\Big|}\;\;\;\;\;\;\;\;
%\end{eqnarray}}
%\textcolor{blue}{Thus we can define a quantifier of non-Markovianity as
\begin{equation}
\label{eq:13_main}
    \mathcal{M}_{NM}=\max_{\rho_{s}(0)}\int_0^{\infty} \overline{M}\big[t,\rho_s(0)\big]dt.
\end{equation}
%\textcolor{red}{\sout{for the dynamics of $m+n$ qubits, $m$ of which are connected to Markovian baths, while $n$ are to non-Markovian ones.}} %\textcolor{blue}{Here $M^{n^+}_{NM}$ is the positive part of $M^{n}_{NM}$.}
%for arbitrary number of qubits coupled to arbitrary numbers of Markovian and non-Markovian baths. 
For a Markovian dynamics, %\textcolor{red}{\sout{the local superoperators as well as the total GKSL equation guarantees the existence of steady states, and therefore the quantifier appears with $\overline{M}_{NM}^0=0$}} 
we get $\mathcal{M}_{NM}=0$. In case there is at least one non-Markovian bath, the quantifier 
%\textcolor{red}{\sout{$\overline{M}_{NM}^n$}} 
$\mathcal{M}_{NM}$ may give 
%\textcolor{red}{\sout{gives}} 
a positive (non-zero) value. For a single system, $\mathcal{M}_{NM}$ reduces to the well-known BLP measure of non-Markovianity proposed in Ref.~\cite{Piilo} within some restrictions, while for higher number of parties this equality does not hold. See Appendix~\ref{appen3} in this regard. Note that the quantifiers of non-Markovianity described in the literature are typically not easily computable. The quantifier %\textcolor{red}{\sout{$\overline{M}_{NM}^n$}} 
$\mathcal{M}_{NM}$ is, however, easily computable, and therein lies its potential utility, viz. in providing a computable strength of non-Markovianity in the dynamics of a system. In Appendix~\ref{appen4}, we have explored how introducing non-Markovian baths or substituting Markovian baths with non-Markovian ones impacts the dynamics of the system. We find that, initially non-Markovian baths have a strong effect, but over time, Markovian baths dominate, suppressing the amplitude of oscillations of the witness of non-Markovianity 
to approximately zero. This is expected because for
a long time, the impact of memory effects, arising from the presence of non-Markovianity, diminishes. Also, more the number of Markovian baths, the quicker is the suppression. For a complete non-Markovian situation, where all the baths are from the non-Markovian family, the periodic oscillatory pattern of the witness of non-Markovianity %\textcolor{red}{\sout{the quantifier}} 
gets disrupted. 
%\textcolor{red}{\sout{We have studied in the Appendix, how the non-Markovianity of the dynamics increases by adding a non-Markovian bath, or by replacing a Markovian bath with a non-Markovian one. %it is necessary to study the properties of the quantifier, specially how the quantifier behaves in time with the increase of the non-Markovianity of the evolution.
%In the next subsection we depict the time dynamics of $\overline{M}_{NM}^n$ for increasing strength of non-Markovianity.
%We find that while non-Markovian baths have a significant contribution for times near the initial times, for larger time, the effect of Markovian baths dominate, suppressing the amplitude of oscillations of the \textcolor{blue}{witness of} non-Markovianity 
%\textcolor{red}{\sout{quantifier}} 
%to approximately zero and, more the number of Markovian baths, the quicker is the suppression.}}
\par
\noindent
\section{Conclusion}
%\emph{\textbf{Conclusion.--}}
\label{sec:5}
To summarize, 
%\textcolor{blue}{we have derived a GKSL-like equation for scenarios involving multiple subsystems, each interacting locally with separate heat baths – some being Markovian and others non-Markovian.} 
we have derived the GKSL-like equation for a situation containing more than one system, each interacting locally with a separate heat bath, some of which are Markovian, while others are non-Markovian. We present the dynamics of a multipartite system evolving under a mixture of Markovian and non-Markovian local environments.
%has, to our knowledge,  not been studied previously. 
Our work provides a significant broadening of the area of investigation of open quantum dynamics, 
%\textcolor{blue}{ relevant to real-world settings, including hybrid systems like atom-photon setups.} 
as a combination of non-Markovian and Markovian environments is a reasonable possibility in a realistic situation, especially when considering hybrid physical systems such as atom-photon arrangements.
%Considering a mixed set of baths 
%\textcolor{blue}{Our approach alters the Spohn's theorem to the multi-party context, considering sets of local Markovian and non-Markovian environments. As a consequence of the modification, we obtained a computable quantifier of non-Markovianity, which detects deviations from a Markovian scenario. 
%The computability of our measure sets it apart from many existing complex quantifiers.
%}
Our setup
leads to a modification of the Spohn's theorem, taken to the multiparty case with a 
%\textcolor{red}{\sout{mixed}} 
set of local Markovian and non-Markovian environments. As a consequence of the modification, we obtained a computable quantifier of non-Markovianity, which can detect the deviation from a Markovian situation. Most of the known quantifiers of non-Markovianity available in the literature are not easily computable. The computability of our measure can potentially be
%On the contrary, the quantifier that we have proposed is easily computable, which can be considered as 
an useful tool to detect non-Markovianity. Analysis of the time dynamics of the quantifier expectedly showed that for an evolution affected by a combination of local Markovian and non-Markovian baths, non-Markovian effects are prominent for times close to initial time, but with the increase of time, non-Markovianity of the dynamics decreases and the evolution tends to be more and more Markovian-like.\par
%\textcolor{red}{An important point to be kept in mind that, the quantifier, $\overline{M}_{NM}^n$, although computable, has some limitations in quantifing the non-Markovianity of a dynamics. As in the usual measures and quantifier used in the literature, it cannot evaluate the non-Markovianity of a dynamics with some numerical value. Moreover, it has no precise numerical bound to quantify the non-Markovianity. It can only detect the strength of non-Markovianity by showing an oscillation with a suppressing amplitude. How long the oscillation survives before reaching the Markovian limit is the measure of the strength of non-Markovianity of the evolution. In this manner, in our search we can infer $\overline{M}_{NM}^1 < \overline{M}_{NM}^2 < \overline{M}_{NM}^3< \overline{M}_{NM}^4$. Therefore, we conclude, $\overline{M}_{NM}^n$ acts as a quantifier of non-Markovianity, which measures the strengh of non-Markovianity in case of combined local Markovian and non-Markovian dynamics.}
\noindent
\section{Acknowledgments}
%\emph{\textbf{Acknowledgements.--}}
We acknowledge computations performed using Armadillo~\cite{Sanderson,  Sanderson1} and QIClib~\cite{QIClib} on the cluster computing facility of the Harish-Chandra Research Institute, India. This research was supported in part by the `INFOSYS scholarship for senior students'.  We also acknowledge partial support from the Department of Science and Technology, Government of India, through QuEST with Grant No. DST/ICPS/QUST/Theme-3/2019/120.

\appendix

\section{Derivation of the two-party GKSL-like equation for local Markovian and non-Markovian baths}
\label{appen1}
\begin{figure*} 
\includegraphics[height=6cm,width=8.5cm]{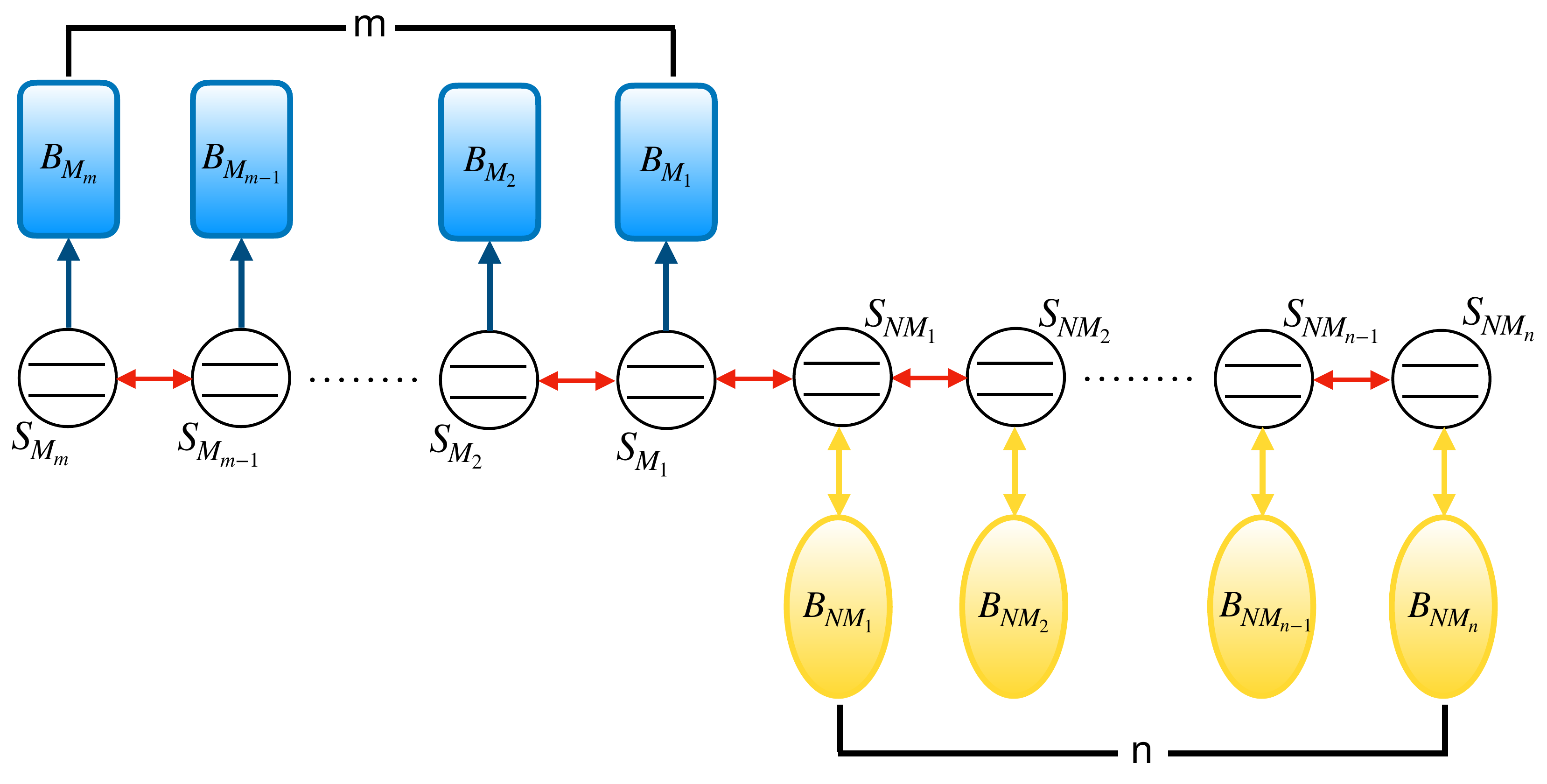}
\includegraphics[height=6cm,width=8cm]{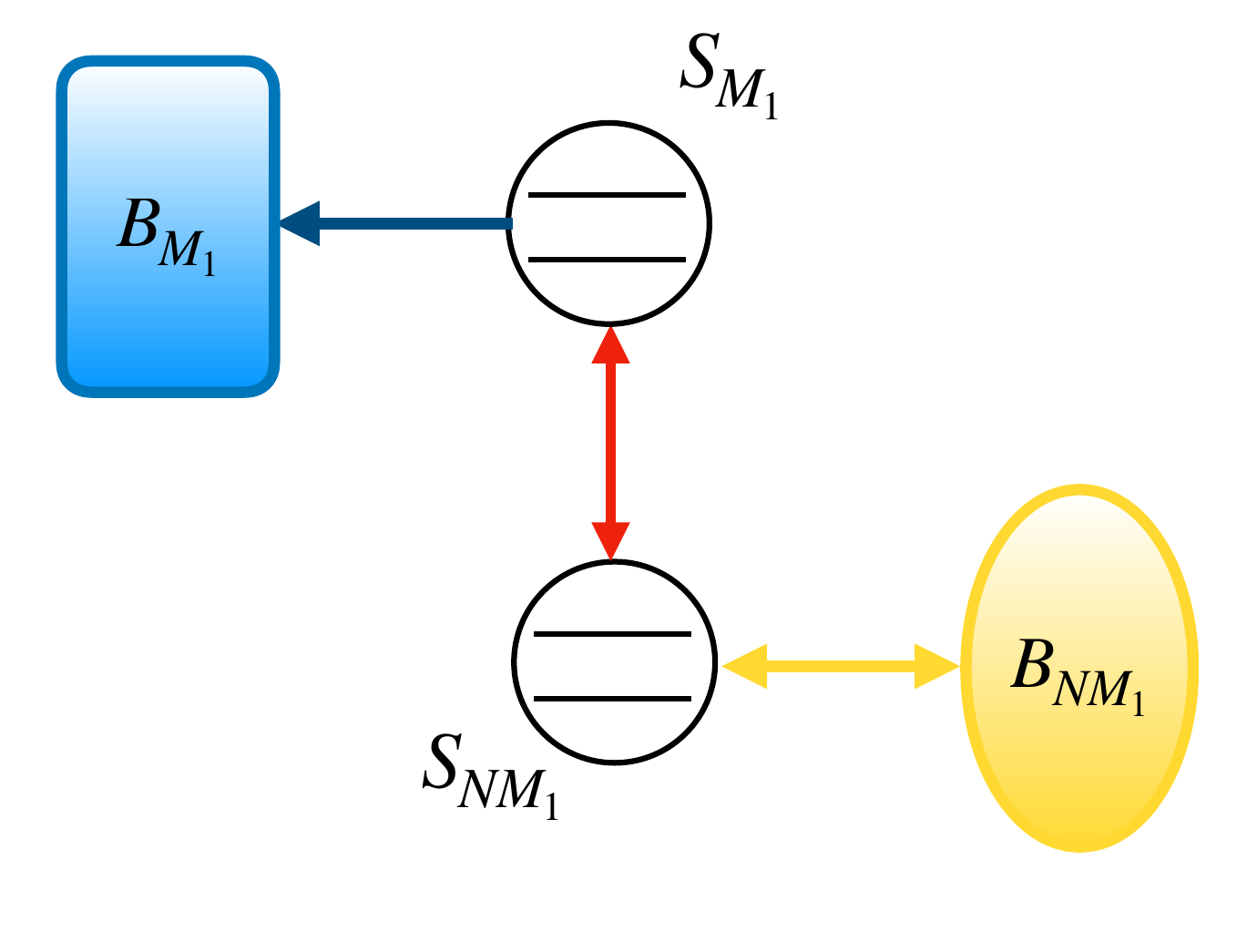}
\caption{A combination of local environments. In the left panel, we present a schematic diagram of $m+n$ TLSs, evolving under their system Hamiltonian as well as local environments, which are some Markovian and the rest not so. $B_{M_1} \cdots B_{M_m}$ are the baths, which can be treated under the Born-Markov approximations, hence are Markovian baths, and $B_{N_1} \cdots B_{N_n}$ are the baths residing in a non-Markovian family. A special case of the left panel is presented in the right one where only two TLSs are interacting with two baths locally among which one is Markovian and the other is non-Markovian. $S_{M_1}$ and $S_{NM_1}$ are the two TLSs. $B_{M_1}$ is the Markovian bath, while $B_{NM_1}$ is non-Markovian one.}
%\figuretag{S1}
\label{fig:S1_appen}
\end{figure*}
A schematic diagram of the two-party two-bath setup is depicted in the right panel of Fig.~\ref{fig:S1_appen}. (The left panel depicts the more general case of an arbitrary number of parties with some connected to  Markovian baths and the rest to non-Markovian ones.)
The Hamiltonian of the composite setup is given by $H=H_s+H_B+H_I$, where $H_s$ describes the Hamiltonian of the composite system consisting of the two parties, $H_B$ stands for the combined local Hamiltonian of the two baths and $H_I=\sum_X H_{I_{X_1}}$ for $X \in \{M,NM\}$. Here $H_{I_{M_1}}$ represents the interaction between $S_{M_1}$ and $B_{M_1}$, and $H_{I_{NM_1}}$ presents the interaction between $S_{NM_1}$ and $B_{NM_1}$. Note that, the Hamiltonian $H_s$, describing the Hamiltonian of the composite system containing two subsystems, is a general Hamiltonian encompassing both the local and interacting part of the two subsystems. Precisely, this $H_s$ can be written as $H_s=H_{\text{loc}}+V$, where $H_{\text{loc}}$ is the local Hamiltonian of the two subsystems and $V$ represents the interaction between them. In the Schr\"odinger picture, let the density matrix of the composite two-party two-bath setup at time $t$ be represented by $\rho(t)$. It is useful to perform the calculation in the interaction picture~\cite{Petruccione}.
The von Neumann equation in interaction picture will be
\begin{equation}
\label{eq:liouville_appen}
    \frac{d\rho_{\text{int}}(t)}{dt}=-\frac{i}{\hbar}\Big[H_{I_{\text{int}}}(t),\rho_{\text{int}}(t)\Big],
    \tag{A1}
\end{equation}
where $H_{I_{\text{int}}}(t)=e^{\frac{i(H_s+H_B)t}{\hbar}}H_Ie^{-\frac{i(H_s+H_B)t}{\hbar}}$, $\rho_{\text{int}}(t)=e^{\frac{i(H_s+H_B)t}{\hbar}}\rho(t)e^{-\frac{i(H_s+H_B)t}{\hbar}}$, 
%\begin{eqnarray}
%    &&H_{I_{int}}(t)=e^{\frac{i(H_s+H_B)t}{\hbar}}H_Ie^{-\frac{i(H_s+H_B)t}{\hbar}},\nonumber\\
%    &&\rho_{int}(t)=e^{\frac{i(H_s+H_B)t}{\hbar}}\rho(t)e^{-\frac{i(H_s+H_B)t}{\hbar}},
%\end{eqnarray}
without assuming a commutativity relation of $H_I$ and $\rho(t)$ with $H_s$ and $H_B$. The interaction Hamiltonian in the Schr\"odinger picture can be decomposed in the form~\cite{Petruccione,Rivas}, $H_{I_{X_1}}=\sum_{k} A_{{X_1}_k} \otimes B_{{X_1}_k}$, with $A_{{X_1}_k}$ and $B_{{X_1}_k}$ being the system and bath operators respectively. Here $X \in \{M,NM\}$ and $k$ runs from 1 to the maximum number of terms required for the decomposition for each $j$. Reverting to the interaction picture we get, $H_{I_{{X_1},\text{int}}}(t)= \sum_{k} A_{{X_1}_{k}}(t) \otimes B_{{X_1}_{k}}(t)$, where $A_{{X_1}_{k}}(t)= e^{\frac{iH_st}{\hbar}} A_{{X_1}_{k}} e^{-\frac{iH_st}{\hbar}}$, $B_{{X_1}_{k}}(t)=e^{\frac{iH_Bt}{\hbar}} B_{{X_1}_{k}} e^{-\frac{iH_Bt}{\hbar}}$. 
%Here $i=1,2$ and $k=\alpha,\beta$. 
The system operators $A_{{X_1}_{k}}$ can be decomposed in the eigenspace of the system Hamiltonian as $A_{{X_1}_{k}}(\omega)=\sum_{\epsilon^{\prime}-\epsilon=\omega}\Pi(\epsilon)A_{{X_1}_{k}} \Pi(\epsilon^{\prime}),$ where $\Pi(\epsilon)$'s are the projectors onto the eigenspaces of the system Hamiltonian $H_s$ corresponding to the eigenfrequency $\epsilon$. Therefore, we have the properties, $[H_s,A_{{X_1}_{k}}(\omega)]=-\hbar\omega A_{{X_1}_{k}}(\omega)$ and $[H_s,A_{{X_1}_{k}}^{\dagger}(\omega)]=\hbar\omega A_{{X_1}_{k}}^{\dagger}(\omega)$, indicating the fact that $A_{{X_1}_{k}}(t)$ and $A_{{X_1}_{k}}(\omega)$ are related by a Fourier transformation from the $t$ space to the $\omega$ space, as $A_{{X_1}_{k}}(t)=\sum_\omega e^{-i\omega t}A_{{X_1}_{k}}(\omega)$ and $A_{{X_1}_{k}}^{\dagger}(t)=\sum_\omega e^{i\omega t}A_{{X_1}_{k}}^{\dagger}(\omega)$. Now the interaction Hamiltonian becomes
\begin{equation}
     H_{I_{\text{int}}}(t)=\sum_X H_{I_{{X_1},\text{int}}}= \sum_{X,k,\omega} e^{-i\omega t} A_{{X_1}_k}(\omega) \otimes B_{{X_1}_k}(t). %+\sum_{\beta,\omega} e^{-i\omega t} A_{2_{\beta}}(\omega) \otimes B_{2_{\beta}}(t).
     \tag{A2}
\end{equation}
The state of the entire setup $\rho_{\text{int}}(t)$, is given by
\begin{equation}
\label{eq:integral_appen}
    \rho_{\text{int}}(t)=\rho(0)-\frac{i}{\hbar}\int_0^t ds [H_{I_{\text{int}}}(s),\rho_{\text{int}}(s)].
    \tag{A3}
\end{equation}
We assume that initially the systems are uncorrelated to the baths, and that the baths themselves are also in a product state, so that at time $t=0$, $\rho(0)=\rho_s(0)\otimes \rho_{B_{M_1}}(0) \otimes \rho_{B_{NM_1}}(0)$.
%\begin{equation}
%    \rho(0)=\rho_s(0)\otimes \rho_{B_1}(0) \otimes \rho_{B_2}(0),
%\end{equation}
%\textcolor{red}{\sout{, where \textcolor{blue}{$B_{M_1}$ and $B_{NM_1}$} are initially in their canonical equilibrium states given by \textcolor{blue}{$\rho_{B_{X}}(0)=\frac{e^{-\beta_{X}H_{B_X}}}{\text{tr}(e^{-\beta_{X}H_{B_X}})}$ for $X \in \{M_1,NM_1\}$}. The corresponding \textcolor{blue}{$\beta_X$'s are $\frac{1}{k_BT_X}$}, with $k_B$ being the Boltzmann constant.}}
As we mentioned earlier, that the bath $B_{M_1}$ is Markovian and therefore in the further calculations, while talking about $B_{M_1}$, we will impose the Born-Markov approximations, which tells that the coupling between the subsystem $S_{M_1}$ and $B_{M_1}$ is weak, so that the state of $B_{M_1}$ regains its initial state after every time step of interaction with $S_{M_1}$, and that any correlation created between $B_{M_1}$ and $S_{M_1}$ is also destroyed after the same time step. Moreover, $B_{M_1}$ will also be assumed to remain uncorrelated with $B_{NM_1}$ during the evolution. So, at time $t$, the state of the entire setup will take the form, $\rho(t) \approx \rho_{sB_{NM_1}}(t) \otimes \rho_{B_{M_1}}$,
%\begin{equation}
%    \rho(t) \approx \rho_{sB_2}(t) \otimes \rho_{B_1},
%    \end{equation}   
where $\rho_{sB_{NM_1}}(t)$ is the density matrix of the systems $S_{M_1}$ and $S_{NM_1}$ combined along the bath $B_{NM_1}$ at time $t$, and $\rho_{B_{M_1}}=\rho_{B_{M_1}}(0)$. Now we make a further assumption for the Markovian bath $B_{M_1}$~\cite{Petruccione}, viz.
\begin{equation}
\label{eq:ini_assump_appen}
    \text{tr}_B[H_{I_{M_1,\text{int}}}(t),\rho_{\text{int}}(0)]=0.
    \tag{A4}
\end{equation}
As a consequence of this assumption, the Markovian bath $B_{M_1}$ possesses the property,
\begin{equation}
\label{eq:B1_appen}
    \text{tr}_{B_{M_1}}(B_{{M_1}_k}\rho_{B_{M_1}})=0.
    \tag{A5}
\end{equation} 
Here we assume $B_{M_1}$ is initially in its stationary state, i.e., $[H_{B_{M_1}},\rho_{B_{M_1}}(0)]=0$. This Eq.~(\ref{eq:B1_appen}) 
%\textcolor{red}{\sout{which}}
is a very important relation for the succeeding calculations of this paper. %\textcolor{red}{\sout{our paper}}. 
Now, using the integral form of $\rho_{\text{int}}(t)$, given in Eq.~(\ref{eq:integral_appen}), and then using Eq.~(\ref{eq:ini_assump_appen}) in the von Neumann equation, we get
\begin{align}
   &\frac{d\rho_{\text{int}}(t)}{dt}=\nonumber\\
   &\sum_X -\frac{1}{\hbar^2}\int_0^t ds \Big[H_{I_{{X_1},\text{int}}}(t),\big[H_{I_{{X_1},\text{int}}}(s),\rho_{sB_{NM_1},\text{int}}(s) \nonumber\\
   %&&-\frac{1}{\hbar}\int_0^t ds \Big[H_{I_{1,int}}(t),\big[H_{I_{2,int}}(s),\rho_{sB_2,int}(s) \otimes \rho_{B_1}\big]\Big]\nonumber\\
   &\phantom{a}\otimes \rho_{B_{M_1}}\big]\Big]-\frac{i}{\hbar}\Big[H_{I_{NM_1,\text{int}}}(t),\rho_{sB_{NM_1},\text{int}}(t) \otimes \rho_{B_{M_1}}\Big],
   \tag{A6}
\end{align}
for $X \in \{M,NM\}$. The first term for $X=NM$ vanishes by using the relation given in Eq.~(\ref{eq:B1_appen}). 
%So the dynamics of the system,
%\begin{eqnarray}
%    \frac{d\rho_{int}(t)}{dt}&=&-\frac{1}{\hbar}\int_0^t ds \Big[H_{I_{1,int}}(t),\big[H_{I_{1,int}}(s),\rho_{sB_2,int}(s) \otimes \rho_{B_1}\big]\Big]\nonumber\\
%   &&-\frac{i}{\hbar}\Big[H_{I_{2,int}}(t),\rho_{sB_2,int}(t) \otimes \rho_{B_1}\Big].
%\end{eqnarray}
Thus the reduced system dynamics comes out to be 
%Eq.~(S1) in the Appendix.
\begin{align}
    &\frac{d\tilde{\rho}_{s,\text{int}}(t)}{dt}=\nonumber\\
    &-\frac{1}{\hbar^2}\int_0^t ds\; \text{tr}_{B_{M_1}}\Big[H_{I_{M_1,\text{int}}}(t),\big[H_{I_{M_1,\text{int}}}(s),\tilde{\rho}_{s,\text{int}}(s) \nonumber\\
    &\phantom{a}\otimes \rho_{B_{M_1}}\big]\Big]-\frac{i}{\hbar}\text{tr}_B\Big[H_{I_{NM_1,\text{int}}}(t),\rho_{sB_{NM_1},\text{int}}(t) \otimes \rho_{B_{M_1}}\Big].
\tag{A7}
 \end{align}
Here $\tilde{\rho}_{s,\text{int}}(t)=\text{tr}_{B_{NM_1}}\{\rho_{sB_{NM_1},\text{int}}(t)\}$. 
%\textcolor{red}{
Next we use the Markov approximation, i.e., replace the integrand $\tilde{\rho}_{s,\text{int}}(s)$ in the first term by $\tilde{\rho}_{s,\text{int}}(t)$, 
%replace the integrand $\tilde{\rho}_{s,int}(s)$ in the Markovian part of the RHS of (S1)
%first term 
%by $\tilde{\rho}_{s,int}(t)$
so that the development of the reduced state of the system at time $t$ only depends on the present state $\tilde{\rho}_{s,\text{int}}(t)$.
%, i.e., the system has no memory of the evolution.} 
Furthermore, we substitute $s$ by $t-s$ and let the upper limit of the integral go to infinity so that we can avoid the dependency of the reduced density matrix on the explicit choice of the initial preparation at time $t=0$. See Ref.~\cite{Petruccione} for more details about the Markovian approximations. Hence, after
%\begin{eqnarray}
%    \frac{d\tilde{\rho}_{s,int}(t)}{dt}&=&-\frac{1}{\hbar}\int_0^t ds \text{tr}_{B_1}\Big[H_{I_{1,int}}(t),\big[H_{I_{1,int}}(s),\tilde{\rho}_{s,int}(t) \otimes \rho_{B_1}\big]\Big]\nonumber\\
%   &&-\frac{i}{\hbar}\text{tr}_B\Big[H_{I_{2,int}}(t),\rho_{sB_2,int}(t) \otimes \rho_{B_1}\Big].
%\end{eqnarray}
% So . Thus
%\begin{eqnarray}
%    &&\frac{d\tilde{\rho}_{s,int}(t)}{dt}=\nonumber\\
%    &&-\frac{1}{\hbar^2}\int_0^\infty ds\; \text{tr}_{B_1}\Big[H_{I_{1,int}}(t),\big[H_{I_{1,int}}(t-s),\tilde{\rho}_{s,int}(t) \otimes \rho_{B_1}\big]\Big]\nonumber\\
%   &&\phantom{abar jonmo nebo}-\frac{i}{\hbar}\text{tr}_B\Big[H_{I_{2,int}}(t),\rho_{sB_2,int}(t) \otimes \rho_{B_1}\Big].
%\end{eqnarray}
 %  &=&\frac{1}{\hbar}\int_0^\infty ds \text{tr}_{B_1}\Big[H_{I_{1,int}}(t)\tilde{\rho}_{s,int}(t) \otimes \rho_{B_1}H_{I_{1,int}}(t-s)\nonumber\\
%   &&-H_{I_{1,int}}(t)H_{I_{1,int}}(t-s)\tilde{\rho}_{s,int}(t) \otimes \rho_{B_1}\Big]+\text{h.c.}\nonumber\\
%   &&-\frac{i}{\hbar}\text{tr}_B\Big[H_{I_{2,int}}(t),\rho_{sB_2,int}(t) \otimes \rho_{B_1}\Big]\nonumber\\
applying the rotating wave approximation, the right-hand side of the above equation comes out to be
\begin{align}
\label{appen:A8}
   &=\frac{1}{\hbar^2}\sum_{\omega}\sum_{k,l} \gamma_{k,l}(\omega) \Big[A_{{M_1}_l}(\omega)\tilde{\rho}_{s,\text{int}}(t)A_{{M_1}_k}^{\dagger}(\omega)\nonumber\\
   &\phantom{aj aloy aloy}-A_{{M_1}_k}^{\dagger}(\omega)A_{{M_1}_l}(\omega)\tilde{\rho}_{s,\text{int}}(t)\Big]+\text{H.c.}\nonumber\\
   &\phantom{jaoyar pother}-\frac{i}{\hbar}\text{tr}_{B_{NM_1}}\Big[H_{I_{{NM_1},\text{int}}}(t),\rho_{sB_{NM_1},\text{int}}(t)\Big].
   \tag{A8}
 \end{align}
$\gamma_{k,l}(\omega)$ is given by $\gamma_{k,l}(\omega)=\int_0^\infty ds e^{i\omega s} \text{tr}_{B_{M_1}}\{B_{M_{1_k}}^{\dagger}(t)B_{M_{1_l}}(t-s)\rho_{B_{M_1}}\}$.
% \begin{equation}
 %    \gamma_{k,l}(\omega)=\int_0^\infty ds e^{i\omega s} \text{tr}_{B_1}\{B_k^{\dagger}(t)B_l(t-s)\rho_{B_1}\}.
 %\end{equation}
 Now we go to the Schr\"{o}dinger picture. $\rho_{B_{M_1}}$ will be the same in both the pictures as $\rho_{B_{M_1}}$ commutes with $H_{B_{M_1}}$. So, $\tilde{\rho}_{s,\text{int}}(t)=\text{tr}_{B_{NM_1}}\Big[e^{\frac{i\big(H_s+H_{B_{NM_1}}\big)t}{\hbar}}\rho_{sB_{NM_1}}(t)e^{-\frac{i\big(H_s+H_{B_{NM_1}}\big)t}{\hbar}}\Big]$.
%(S1) comes out to be as in (S2) in the Appendix.
%
% \begin{equation}
%     \tilde{\rho}_{s,int}(t)=\text{tr}_{B_2}\Big[e^{\frac{i(H_s+H_{B_2})t}{\hbar}}\rho_{sB_2}(t)e^{-\frac{i(H_s+H_{B_2})t}{\hbar}}\Big].
% \end{equation}
Using $\tilde{\rho}_s(t)=\text{tr}_{B_{NM_1}}\big\{e^{\frac{iH_{B_{NM_1}}t}{\hbar}}\rho_{sB_{NM_1}}(t)e^{\frac{-iH_{B_{NM_1}}t}{\hbar}}\big\}=\text{tr}_{B_{NM_1}}(\rho_{sB_{NM_1}})$, the reduced system dynamics in the Schr\"odinger picture turns out to be
% \begin{eqnarray}
%     &&\frac{d\tilde{\rho}_{s}(t)}{dt}=\mathcal{L}\big(\tilde{\rho}_{s}(t)\big)\equiv\nonumber\\
%     &&-\frac{i}{\hbar}\Big[H_s,\tilde{\rho}_s(t)\Big]+\frac{1}{\hbar^2}\sum_{\omega}\sum_{k,l} \gamma_{k,l}(\omega) \Big(A_{1_l}(\omega)\tilde{\rho}_{s}(t)A_{1_k}^{\dagger}(\omega)\nonumber\\
 %  &&-\frac{1}{2}\big\{A_{1_k}^{\dagger}(\omega)A_{1_l}(\omega),\tilde{\rho}_{s}(t)\big\}\Big)-\frac{i}{\hbar}\text{tr}_{B_2}\Big[H_{I_{2}}(t),\rho_{sB_2}(t)\Big].\;\;\;\;\;\;\;
% \end{eqnarray}
%Here we have used that $\tilde{\rho}_s(t)=\text{tr}_{B_2}\big\{e^{\frac{iH_{B_2}t}{\hbar}}\rho_{sB_2}(t)e^{\frac{-iH_{B_2}t}{\hbar}}\big\}=\text{tr}_{B_2}(\rho_{sB_2})$. 
%\textcolor{red}{[(19) e use kora hoyechhe.]}
%Thus, we can write
\begin{align}
\label{eq:20_appen}
    \frac{d\tilde{\rho}_{s}(t)}{dt}&=\mathcal{L}\big[\tilde{\rho}_{s}(t)\big]\equiv -\frac{i}{\hbar}\Big[H_s,\tilde{\rho}_s(t)\Big]+\mathcal{D}_{M_1}\big[\tilde{\rho}_{s}(t)\big]\nonumber\\
    &\phantom{ki gabo ami ki shuna}+\mathcal{D}_{NM_1}\big[\rho_{sB_{NM_1}}(t)\big],
    \tag{A9}
\end{align}
where
\begin{align}
    &\mathcal{D}_{M_1}\big[\tilde{\rho}_{s}(t)\big]=\frac{1}{\hbar^2}\sum_{\omega}\sum_{k,l} \gamma_{k,l}(\omega) \Big(A_{M_{1_l}}(\omega)\tilde{\rho}_{s}(t)A_{M_{1_k}}^{\dagger}(\omega)\nonumber\\
   &\phantom{amay akhash bollo}-\frac{1}{2}\big\{A_{M_{1_k}}^{\dagger}(\omega)A_{M_{1_l}}(\omega),\tilde{\rho}_{s}(t)\big\}\Big),\nonumber\\
   &\mathcal{D}_{NM_1}\big[\rho_{sB_{NM_{1}}}(t)\big]=-\frac{i}{\hbar}\text{tr}_{B_{NM_1}}\Big[H_{I_{NM_{1}}}(t),\rho_{sB_{NM_1}}(t)\Big].\;\;\;\;\;
   \label{eq:21_appen}
   \tag{A10}
\end{align}
\section{Alteration of Spohn's theorem in case of all non-Markovian environments}
\label{appen2}
The modified form of the Spohn's theorem for the combined local evolution under Markovian and non-Markovian environments is presented in Eq.~(\ref{eq:33}) for a two-party setup and the generalization of this two-party scenario to a multiparty situation is given in Eq.~(\ref{NM_n1}). This altered version of the Spohn's theorem has been derived by imposing strict restrictions on the Markovian environments. If the restrictions corresponding to the Markovian baths are relaxed for the Markovian environments, i.e., if we consider all the environments to be non-Markovian, i.e., $m=0$, then there will only be the non-Markovian dissipator $\mathcal{D}_{NM_j}\big[\rho_{sB_{NM_{1 \cdots n}}}(t)\big]$  for $j=1$ to $n$ in the GKSL-like master equation given in Eq.~(\ref{eq:22}) of the main text. Let us first consider the simplest situation of the two-party two-bath setup. If both the baths are non-Markovian, then we will not get the term $-k_B\text{tr}\big\{\mathcal{L}_{M_1}\big[\tilde{\rho}_s(t)\big]\big(\ln(\tilde{\rho}_s(t))-\ln(\tilde{\rho}_{\text{th}_{M_1}})\big)\big\}$ in the RHS of Eq.~(\ref{eq:NM1}) of the main text. Instead, we will get the equation,
\begin{align}
    &\sigma_{\{NM_1,NM_2\}}+\sum_{j=1}^2 k_B\text{tr}\big\{\mathcal{L}_{NM_j}\big[\rho_{sB_{NM_{12}}}(t)\big]\big(\ln(\tilde{\rho}_{s}(t))\nonumber\\
    &\phantom{borne gondhe chhonde gitite}-\ln(\tilde{\rho}_{\text{th}_{NM_j}})\big)\big\}=0.
    \tag{B1}
\end{align}
For $n$ parties, $j$ will run from 1 to $n$ in this equation, $\sigma_{\{NM_1,NM_2\}}$ is to be replaced by $\sigma_{\{NM_1,\cdots,NM_n\}}$, and $\rho_{sB_{NM_{12}}}(t)$ will be replaced by $\rho_{sB_{NM_{1\cdots n}}}(t)$. Therefore, this equation will now become the altered version of the Spohn’s theorem, valid for the situation where all the environments exhibit non-Markovian behavior, and is a direct consequence of the balance equation and the concept of EPR. This version of the Spohn’s result does not incorporate any assumption about weak coupling, or correlation destruction (system-bath as well as bath-bath), or about an evolution being memoryless.
%The positivity of this second term is not guaranteed and in the example with a four-qubit system shown below, we have observed that the term may be positive or negative. Therefore, no altered version of Spohn’s theorem can be constructed for the situation where all the environments exhibit non-Markovian behavior. The modification of Spohn’s theorem is only valid for a combined local evolution under Markovian and non-Markovian environments.

\section{Relation of $\mathcal{M}_{NM}$ with the BLP measure of non-Markovianity}
\label{appen3}
In this section, we aim to establish a relationship between the proposed quantifier of non-Markovianity in the main text, denoted as $\mathcal{M}_{NM}$, and the well-known BLP measure~\cite{Piilo}. First, let us introduce the BLP measure of non-Markovianity. For any two quantum states $\rho_1$ and $\rho_2$, we define the rate of change of the distance between these states over time as
\begin{equation}
    \Lambda(t,\rho_{1,2}(0))=\frac{d}{dt} D(\rho_1(t),\rho_2(t)),
    \tag{C1}
\end{equation}
where $\rho_i(t)=\Phi(t)(\rho_i(0))$ for $i=1$ and $2$, and $\Phi(t)(\cdot)$ represents any quantum channel. The function $D$ denotes a distance measure in the space of quantum states. For the purposes of this discussion, we will choose the relative entropy as the ``distance" measure, which leads to
\begin{equation}
    \Lambda(t,\rho_{1,2}(0))=\frac{d}{dt} S(\rho_1(t),\rho_2(t)).
    \tag{C2}
\end{equation}
When $\Lambda\le 0$, the dynamical process exhibits the divisibility property of a dynamical semigroup, indicating a Markovian dynamics. However, if divisibility breaks during the dynamical evolution of the system, $\Lambda$ may take a positive (non-zero) value. As a result, $\Lambda$ serves as a witness of non-Markovianity. To quantify the non-Markovian behavior, we can therefore define the corresponding measure of non-Markovianity as
\begin{equation}
\mathcal{N}=\max_{\rho_{1,2}(0)}\int_{\Lambda>0} \Lambda(t,\rho_{1,2}(0)) dt,
\tag{C3}
\end{equation}
where $\mathcal{N}=0$ for all Markovian processes, i.e., the ones that satisfy the divisibility property of a quantum dynamical semigroup. If $\mathcal{N}>0$ the evolution must be non-Markovian. This is the quantifier of non-Markovianity proposed by Breuer-Laine-Piilo in Ref.~\cite{Piilo}.

To facilitate a comparison between the non-Markovianity quantifier of BLP and the quantifier $\mathcal{M}_{NM}$ [proposed in Eq.~(\ref{eq:13_main}) of the main text], we will impose an additional restriction on the BLP quantifier. Consider a scenario, where a single system, described by the Hamiltonian $H_s$, is immersed in a heat bath, initially kept in its canonical equilibrium state $\tilde{\rho}_{\text{th}_{B_1}}=\frac{e^{-\beta_1 H_{B_1}}}{\text{tr}(e^{-\beta_1 H_{B_1}})}$ with the inverse temperature $k_B\beta_1$ and $H_{B_1}$ being the free Hamiltonian of the bath. Consequently, the canonical equilibrium state of the system can be expressed as $\tilde{\rho}_{\text{th}_1}=\frac{e^{-\beta_1 H_{s}}}{\text{tr}(e^{-\beta_1 H_{s}})}$. Now, we set $\rho_2(t)=\tilde{\rho}_{\text{th}_1}$ for all time. Hence the BLP witness of non-Markovianity $\Lambda$ reduces to
\begin{equation}
    \Lambda(t,\rho_{1}(0),\tilde{\rho}_{\text{th}_1})=\frac{d}{dt} S(\rho_1(t),\tilde{\rho}_{\text{th}_1}).
    \tag{C4}
\end{equation}
Accordingly, the corresponding quantifier of non-Markovianity can be expressed as
\begin{equation}
    \tilde{\mathcal{N}}=\max_{\rho_{1}(0)}\int_{\Lambda>0} \Lambda(t,\rho_{1}(0),\tilde{\rho}_{\text{th}_1}) dt.
    \tag{C5}
\end{equation}

%Let us now consider a single qubit system coupled to an environment which is initially in its thermal equilibrium state. 
In contrast, for the case of a single system, the witness of non-Markovianity, denoted as $\overline{M}\big[t,\rho_s(0)\big]$ in the main text, is given by
\begin{align}
    \overline{M}\big[t,\rho_s(0)\big]&=\max\{0,M^{1}\big[\rho^{\prime}(t)\big]\}\nonumber\\
    &=\max\Big\{0,\frac{d}{dt} S(\tilde{\rho}_s(t)||\tilde{\rho}_{\text{th}_{1}})\Big\},
    \tag{C6}
\end{align}
where $S(\tilde{\rho}_s(t)||\tilde{\rho}_{\text{th}_{1}})=k_B\text{tr}\big\{\mathcal{L}_{1}\big[\rho^{\prime}(t)\big]\big(\ln(\tilde{\rho}_s(t))-\ln(\tilde{\rho}_{\text{th}_{1}})\big)\big\}$. Therefore, the quantifier of non-Markovianity becomes
\begin{align}
    \mathcal{M}_{NM}&=\max_{\rho_{s}(0)}\int_0^{\infty} \overline{M}\big[t,\rho_s(0)\big]dt\nonumber\\
    &=\tilde{\mathcal{N}}.
    \tag{C7}
\end{align}
Hence, we observe that the quantifier of non-Markovianity $\mathcal{M}_{NM}$ reduces to the BLP measure when the state $\rho_2(t)$ in BLP quantifier is fixed at $\tilde{\rho}_{\text{th}_1}$. However, it is important to note that for systems comprising two or more subsystems, this type of relationship between $\mathcal{M}_{NM}$ and $\tilde{\mathcal{N}}$ is not attainable.

\section{Four qubits coupled to a combination of four Markovian and non-Markovian heat baths locally}
\label{appen4}
\begin{figure*}
\centering
\includegraphics[width=8cm]{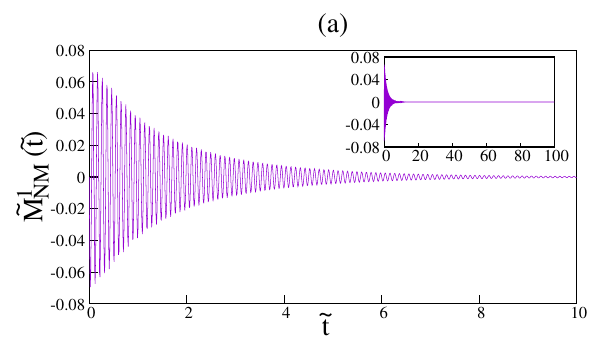}%
\hspace{.25cm}%
\includegraphics[width=8cm]{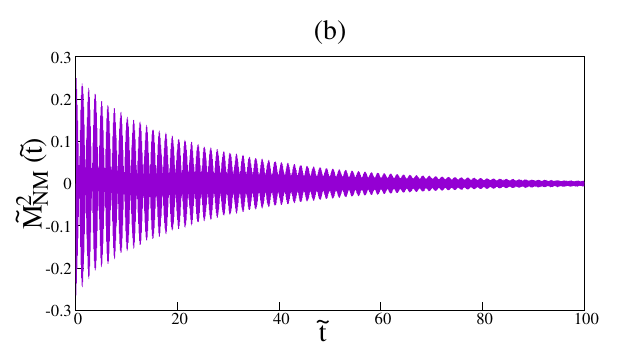}%

\includegraphics[width=8cm]{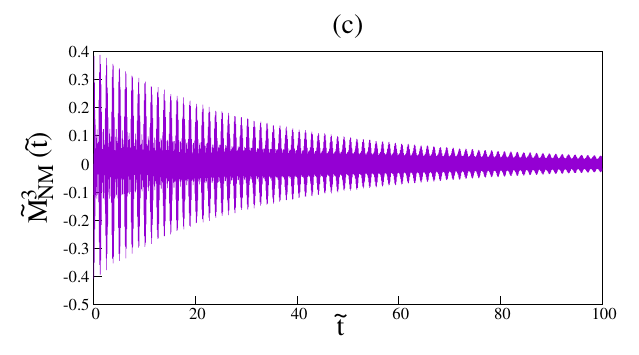}%
\caption{Time dynamics of $\tilde{M}^n_{NM}(\tilde{t})$. %\textcolor{red}{\sout{the non-Markovianity quantifier}}. 
Here we plot the behavior of  $\tilde{M}^n_{NM}(\tilde{t})$ %\textcolor{red}{\sout{the quantifier of non-Markovianity, $\overline{M}_{NM}^n$,}} 
with time for (a) $n=1$, where only the fourth bath is non-Markovian and the other three are Markovian, (b) $n=2$, where the third and fourth baths are non-Markovian and the other two are Markovian, and (c) $n=3$, where only the first bath is Markovian and the other three are non-Markovian, for the initial state of the system taken as 
%\textcolor{blue}{$\rho_s(0)=|\psi_s(0) \rengle \langle \psi_s(0)|$, where} 
$\ket{\psi_s(0)}=\frac{1}{\sqrt{2}}(\ket{0000}+\ket{1111})$. For the demonstration, we have chosen $\omega_1=50.0$, $\omega_2=55.0$, $\omega_3=60.0$, $\omega_4=65.0$, and $\nu_1=\nu_2=\nu_3=\nu_4=1.0$ and the temperatures of the heat baths are set to be $T_1=127.33$, $T_2=105.57$, $T_3=95.8$, and $T_4=68.6$. The coupling constants are taken to be $\kappa_1=\kappa_2=\kappa_3=10^{-3}$ and $\alpha_2=\alpha_3=\alpha_4=0.5$. The quantities plotted along the horizontal axes are dimensionless and the same along the vertical axes have the unit of $k_B \tilde{\eta}$.}
\label{fig:2_appen}
\end{figure*}
We consider here the case of four non-interacting two-level systems (TLSs), each locally immersed in a bath that is either Markovian or not so. To begin, let us consider four non-interacting single-qubit subsystems, each locally immersed in a Markovian bosonic heat bath, kept in their canonical equilibrium states at temperatures $T_1$, $T_2$, $T_3$, and $T_4$, respectively, so that under the time evolution of the system in this circumstance, the 
%all of them are locally immersed in four heat baths, among which all the four are Markovian bosonic baths, for which the 
local canonical equilibrium states are the steady states of the partial superoperators. Here we use the dimensionless temperatures of the baths, defined as $T_j=\frac{\hbar \tilde{\eta}}{k_B \tilde{T_j}}$, where $\tilde{T_j}$ are the real temperatures for $j=1$ to $4$ and $\tilde{\eta}$ is a constant having the dimension of frequency. So, as we discussed in the main text, the quantifier, $\mathcal{M}_{NM}=0$ %\textcolor{red}{\sout{$\overline{M}_{NM}^0=0$}} 
in this case. We now replace the Markovian baths one by one with non-Markovian ones, and study its response in the non-Markovian 
%\textcolor{red}{\sout{quantifier}}
witness $\overline{M}\big[t,\rho_s(0)\big]$ 
%non-Markovian quantifier $\overline{M}_{NM}^n$ 
with time.\par
%in Fig.~\ref{fig:2} for $n=1,2,3$.\par
The Hamiltonian of the composite four-qubit four-bath setup is
\begin{equation}
    H_{tot}=\sum_{j=1}^4 (H_{s_j}+H_{B_{X_j}}+H_{I_{X_j}}),
    \tag{D1}
\end{equation}
with $X$ in the suffixes of the second and the third terms indicating whether the contribution is coming from a Markovian ($M$) or a non-Markovian ($NM$) bath. The local Hamiltonian of each TLS is given by
\begin{equation}
    H_{s_j}=\frac{\hbar\omega_j}{2}\sigma^z_j,
    \tag{D2}
\end{equation}
having the ground-state eigenvalue $-\frac{\hbar\omega_j}{2}$ corresponding to the state $\ket{1}$ and the excited state eigenvalue $\frac{\hbar\omega_j}{2}$ corresponding to the state $\ket{0}$.
For the Markovian harmonic oscillator baths, the local Hamiltonian of each bath can be expressed as
\begin{equation}
    H_{B_{M_j}}=\int_0^{\eta_{max_j}} \hbar \tilde{\eta} d\eta_j a_{\eta_j}^{\dagger} a_{\eta_j},
    \tag{D3}
\end{equation}
and the same for each non-Markovian bath is taken as
\begin{equation}
    H_{B_{NM_j}}=\hbar \nu_j J_{j}^+J_{j}^-.
    \tag{D4}
\end{equation}
For the harmonic oscillator baths,
%\textcolor{red}{\sout{$\tilde{\eta}$ is a constant, being in the unit of frequency,}} 
$a_{\eta_{j}}^{\dagger} (a_{\eta_{j}})$ represents the bosonic creation (annihilation) operator of the harmonic oscillator of the $j^{\text{th}}$ mode of the bath, having the unit of $\frac{1}{\sqrt{\eta_{j}}}$ and connected by the commutation relation $[a_{\eta_{i}},a_{\eta_{i}^\prime}^\dagger]=\delta(\eta_{i}-\eta_{i}^\prime)$, $\eta_{max_j}$ is the cutoff frequency of the $j^{\text{th}}$ Markovian bath. On the contrary, each of the non-Markovian baths, with frequency $\nu_j$ for each $j$, is taken as one described by the spin-star model~\cite{Hutton,Breuer1} consisting of $N$ localized quantum spin-$\frac{1}{2}$ particles centering the single-qubit system at equal distances on a sphere, with $J_j^{\pm}=\sum_{l=1}^N \sigma_{j,(l)}^{\pm}$. Here $\sigma_{j,(l)}^{\pm}=\frac{1}{2}(\sigma^x_{j,(l)}\pm i\sigma^y_{j,(l)})$, with $\vec{\sigma} (\sigma^x,\sigma^y,\sigma^z)$ representing the Pauli matrices. The interaction between the systems and the local bosonic baths is considered as
\begin{equation}
    H_{I_{M_j}}=\int_0^{\eta_{max_j}} \hbar \sqrt{\tilde{\eta}}d\eta_j h(\eta_j) (a_{\eta_j}^{\dagger}\sigma_j^-+a_{\eta_j}\sigma_j^+),
    \tag{D5}
\end{equation}
where $h(\eta_{j})$ is a dimensionless function of $\eta_j$, and represents the system-bath coupling strength. For a harmonic oscillator bath, $\tilde{\eta}h^2(\eta_{j})=\mathcal{J}(\eta_{j})$, where $\mathcal{J}(\eta_{j})$ is the ohmic spectral density function. As we consider the spectral density function as ohmic, $\mathcal{J}(\eta_{j}) \propto \eta_{j}$. Thus $\mathcal{J}(\eta_{j})=\kappa_j \eta_{j}$, where $\kappa_j$ are unitless constants. 
For the qubits connected to non-Markovian spin baths, the interaction is through a Heisenberg $XY$ coupling~\cite{Lieb}, given by
\begin{equation}
    H_{I_{NM_j}}=\hbar\alpha_j(\sigma_{j}^{+}J_{j}^-+\sigma^{-}_{j}J_{j}^+).
    \tag{D6}
\end{equation}
Here $\alpha_j$ quantifies the coupling strength between the $j^{\text{th}}$ system and the $j^{\text{th}}$ non-Markovian bath, having the unit of frequency. The dynamical equation for the $(m+n)=4$-qubit system, locally connected to a combination of Markovian and non-Markovian heat baths, directly follows from Eq.~(\ref{eq:sigma}), and is given by
\begin{align}
    &\frac{d\rho}{d(\tilde{\eta}t)}=\frac{1}{\tilde{\eta}}\mathcal{L}\big[\tilde{\rho}_{s}(t)\big]=-\frac{i}{\hbar\tilde{\eta}}\Big[H_s,\tilde{\rho}_s(t)\Big]\nonumber\\
    &+\frac{1}{\tilde{\eta}}\sum_{j=1}^m\mathcal{D}_{M_j}\big[\tilde{\rho}_{s}(t)\big]+\frac{1}{\tilde{\eta}}\sum_{j=1}^n\mathcal{D}_{NM_j}\big[\rho_{sB_{NM,1\dots n}}(t)\big].
    \tag{D7}
\end{align}
Both  sides of the equation are made dimensionless. For the purpose of our demonstration, we will take the variable \(\tilde{t} = \tilde{\eta}t\) as the dimensionless time.\par
In Fig.~\ref{fig:2_appen}, we have depicted the time dynamics of the 
%\textcolor{red}{\sout{quantifier $\overline{M}_{NM}^n$}}
quantity $\tilde{M}^n_{NM}(\tilde{t})=\sum_{k=1}^{n}M^{k}_{NM}\big[\rho_{sB_{NM_{1\dots n}}}(\tilde{t})\big]$ for $n=1$, $2$, and $3$ in panels (a), (b), and (c), respectively,  $n$ being the number of non-Markovian environments. In all the three panels, $\tilde{M}^n_{NM}(\tilde{t})$ %\textcolor{red}{\sout{the quantifier}} 
exhibits oscillating profiles. The envelope of $\tilde{M}^n_{NM}(\tilde{t})$, having positive and negative values, %\textcolor{red}{\sout{$\overline{M}_{NM}^n$}}
is significant in magnitude for a short time near zero, %reaches a peak and then
but decreases monotonically, tending to zero for large time. The existence of the positive part of this envelope indicates that the witness of non-Markovianity, $\overline{M}\big[\tilde{t},\rho_s(0)\big] > 0$.
%\textcolor{red}{\sout{.This implies that while}} \textcolor{blue}{and} 
We find that the non-Markovian baths have a significant contribution for times near the initial time, while for larger time, the effect of Markovian baths dominate, suppressing the amplitude of oscillations of $\tilde{M}^n_{NM}(\tilde{t})$  %\textcolor{red}{\sout{the non-Markovianity quantifier}} 
to approximately zero. This is expected because for
a long time, the impact of memory effects, arising from the presence of non-Markovianity, diminishes. Although both positive as well as negative oscillations of $\tilde{M}^n_{NM}(t)$ are suppressed in this specific scenario, it is the suppression of the positive oscillations that imply the dominance of the Markovian baths in the evolution. The greater the number of Markovian baths, the quicker is the suppression. [Compare Figs.~\ref{fig:2_appen}(a),~\ref{fig:2_appen}(b), and~\ref{fig:2_appen}(c)]. For a complete non-Markovian situation, where all the baths are from the non-Markovian family, the periodic oscillatory pattern of $\tilde{M}^n_{NM}(\tilde{t})$ %\textcolor{red}{\sout{the quantifier}} 
gets disrupted. The amplitude of oscillation does not diminish with time, as there is no Markovian bath to suppress the oscillation like in case of the combination of local Markovian and non-Markovian dynamics.

\end{document}